\documentclass[aps,prb,twocolumn,amsmath,amsfonts,superscriptaddress]{revtex4-2}
\usepackage{graphicx}
\usepackage{dcolumn}
\usepackage{bm}
\usepackage{boxedminipage}
\usepackage{amsmath}
\usepackage{graphicx}
\usepackage{amsfonts}
\usepackage{feynmf}
\usepackage{mathrsfs}
\usepackage{listings}
\usepackage{algorithm}
\usepackage{algorithmic}
\usepackage{color}
\usepackage{braket}
\usepackage{hyperref,cleveref}
\usepackage{ulem}

\newcommand{\nn}{\nonumber}

\begin{document}

\title{Impurity bands, line-nodes, and anomalous thermal Hall effect\\ in Weyl superconductors}

\author{Taiki Matsushita}
\affiliation{Department of Physics, Graduate School of Science, Kyoto University, Kyoto 606-8502, Japan}
\affiliation{Center for Gravitational Physics and Quantum Information, Yukawa Institute for Theoretical Physics, Kyoto University, Kyoto 606-8502, Japan}
\author{Naoyuki Kimura}
\affiliation{Department of Materials Engineering Science, Osaka University, Toyonaka, Osaka 560-8531, Japan}
\author{Takeshi Mizushima}
\affiliation{Department of Materials Engineering Science, Osaka University, Toyonaka, Osaka 560-8531, Japan}
\author{Ilya Vekhter}
\affiliation{Department of Physics and Astronomy, Louisiana State University, Baton Rouge, LA 70803-4001}
\author{Satoshi Fujimoto}
\affiliation{Department of Materials Engineering Science, Osaka University, Toyonaka, Osaka 560-8531, Japan}
\affiliation{Center for Quantum Information and Quantum Biology, Osaka University, Toyonaka, Osaka 560-8531, Japan}

\date{\today}

\begin{abstract}
We investigate the anomalous thermal Hall effect (ATHE) in Weyl superconductors realized by the $E_{1u}$ ($p$-wave and $f$-wave) chiral superconducting order for the point group $D_{6h}$.
Using the quasiclassical transport theory, we analyze the influence of the impurity scatterings and the line-nodal excitations on the ATHE.
We compare the extrinsic (impurity-induced) ATHE with the intrinsic (topological) ATHE to identify the dominant contribution.
Because the transverse response is sensitive to the slope in the density of states (DOS) at the Fermi energy, the extrinsic ATHE vanishes in both the Born (weak impurity potential) and unitarity (strong impurity potential) limits.
The amplitude of the impurity contribution to the thermal Hall conductivity (THC) reaches maximum between these limits when the slope of the DOS becomes large due to impurity bands near the Fermi energy.
In such situations, the extrinsic ATHE dominates the intrinsic ATHE even at low temperatures.
The extrinsic ATHE is sensitive to line-nodal excitations, whereas the intrinsic ATHE is insensitive to bulk excitations.
When line nodes involve the sign change of the order parameter, the impurity contribution to the THC is suppressed even though the phase space for low-energy excitation is large.
In contrast, if line nodes are not accompanied by such sign changes, the extrinsic ATHE is significantly enhanced.
Our results form a basis for the comprehensive analysis of anomalous thermal transport in Weyl superconductors.
\end{abstract}

\maketitle

\section{Introduction}
\label{Sec1}
Weyl superconductors (WSCs) are time-reversal symmetry broken (TRSB) superconductors whose low-energy excitations behave as Weyl quasiparticles~\cite{volovic}.
They are realized when the superconducting condensate consists of Cooper pairs with a fixed orbital angular momentum and are described by a complex order parameter, $\Delta(\bm k) \propto (k_x\pm ik_y)^\nu \;(\nu \in \mathbb{Z})$.
On a three-dimensional Fermi surface centered on the $\Gamma$ point, the gap closing points at $k_x=k_y=0$ become Weyl points, which are sources and drains of Berry flux~\cite{RevModPhys.82.1959,hosur2013recent,yan2017topological}.
As a result, the near-nodal Bogoliubov quasiparticles behave as Weyl particles.
It is common to both refer to such order parameters and label the corresponding ground states as chiral~\cite{kallin2016chiral}.
Generically, chiral superconductors with a three-dimensional Fermi surface are good candidates for the realization of WSCs.

Among the phenomena that experimentally identify WSCs are the anomalous thermal Hall effect (ATHE), the transverse thermal current driven by a temperature gradient without an applied magnetic field, and the chiral anomaly-induced phenomena, such as the torsional chiral magnetic effect and the negative thermal magnetoresistivity by textures of the order parameters~\cite{read2000paired,sumiyoshi_ATHE,balatskii1986chiral,matsushita2018charge,kobayashi2018negative,ishihara2019torsional,nissinen2020emergent,nissinen2020thermal}.
In particular, the observation of the ATHE unambiguously identifies the chiral ground states.

At the microscopic level, there are two sources for a finite thermal Hall signal in chiral superconductors: intrinsic and extrinsic.
The intrinsic mechanisms arise from the effective Lorentz force generated by the Berry curvature and are characterized by the configuration of Weyl points in momentum space~\cite{meng2012weyl,goswami2013topological,goswami2015topological,nakai2017laughlin,nakai2020weyl,moriya}.
The extrinsic mechanism relies on impurities and the transfer of the angular momentum between the condensate and the quasiparticles during scattering events~\cite{Arfi1988, Spivak:2015,yip2016low,yilmaz2020spontaneous,ngampruetikorn2020impurity}, which contributes both to the energy-dependent skew scattering and the Andreev (inter-branch, electrons to holes and vice versa) scattering at impurity sites.
The skew scattering directly couples to the temperature gradient for $|\nu|=1$ and results in the ATHE signal~\cite{Arfi1988, Spivak:2015,yip2016low}.
The Andreev mechanism only couples to impurities if the impurity potential is non-$s$-wave, with scattering matrix elements coupling different angular-momentum channels, and therefore appears for finite size impurities, dominating the response for $|\nu|\geq 2$~\cite{yilmaz2020spontaneous,ngampruetikorn2020impurity,wave24}.

In this paper, we address two aspects of the ATHE in chiral superconductors with point-like impurities. 
First, we elucidate its origin and demonstrate the relation between the extrinsic ATHE and the evolution of the density of states (DOS) in the sub-gap impurity band, which arises from the broadening of impurity resonant states~\cite{hirschfeld1988consequences, RevModPhys.78.373}.
In particular, the particle-hole anisotropy, which is necessary for transverse transport, sensitively depends on the scattering phase-shift at individual impurities~\cite{PhysRevLett.128.097001,matsushita2024spin}.
Second, we focus on the magnitude of the thermal Hall conductivity (THC) in the situation where the winding number, $\nu$, differs from the total angular momentum of the Cooper pairs, $l$. 
Such a situation often occurs when the chiral superconducting order involves additional nodal excitations.

Our results are relevant to several candidates of WSCs.
Historically, the Anderson-Brinkmann-Morel (ABM) state in $^3$He was the first well-established Weyl superfluid with the chiral $p$-wave pairing~\cite{leggett1975theoretical,volovic,sil14,mizushima2016symmetry,ikegamiscience,ikegamijpsj,shevtov}.
In superconducting materials, the chiral ground states over three-dimensional Fermi surfaces were proposed in a number of materials, including URu$_2$Si$_2$, UPt$_3$, U$_{1-x}$Th$_x$Be$_{13}$, and SrPtAs~\cite{kasahara2009superconducting,kittaka2016evidence,schemm2015evidence,goswami2013topological,yamashita2015colossal,sumiyoshi2014giant,stewart1984possibility,sauls1994order,schemm2014observation,izawa2014pairing,tsutsumi2013upt3,tsutsumi2012spin,yanase2016nonsymmorphic,goswami2015topological,ott1983u,shimizu2017quasiparticle,golding1985observation, PhysRevLett.55.1319,mizushima2018topology,machida2018spin, PhysRevB.87.180503, PhysRevB.89.020509}.
The chiral $d$-wave pairing, $\Delta(\bm k)\propto k_z(k_x\pm ik_y)$, was proposed in URu$_2$Si$_2$ and explains the giant Nernst effect above the critical temperature due to preformed chiral Cooper pairs~\cite{yamashita2015colossal,sumiyoshi2014giant}.
Uranium compounds UPt$_3$ and U$_{1-x}$Th$_x$Be$_{13}$ show spin-triplet superconductivity and multiple superconducting phases as a function of tuning parameters~\cite{adenwalla1990phase,hasselbach1989critical,ott1985phase,kim1991investigation,rauchschwalbe1987phase}.
While the symmetry of the superconducting order parameter is still a subject of debate in these materials, TRSB was observed in the so-called B-phase, suggesting a chiral nature of the condensate~\cite{schemm2014observation, PhysRevLett.65.2816}.
In general, ferromagnetic superconductors show a complex ``nonunitary'' order parameter and support Weyl nodes on three-dimensional Fermi surfaces~\cite{mac01, PhysRevB.66.134504, PhysRevB.90.064506,sau2012topologically, matsushita2024spin}.
Consequently, ferromagnetic UCoGe, URhGe, and UGe$_2$ are also candidates of WSCs~\cite{PhysRevLett.99.067006,mineev2017phase,aoki2001coexistence,saxena2000superconductivity}.

We employ the quasiclassical transport (Eilenberger) theory, which is a hierarchical expansion in $(k_{\rm F}\xi_0)^{-1}\sim T_{\rm c}/T_{\rm F}\ll 1$, where $k_{\rm F}$ is the Fermi momentum, $\xi_0=v_{\rm F}/(2\pi T_{\rm c})$ is the superconducting coherence length, $v_{\rm F}$ is the Fermi velocity, and $T_{\rm c}$ and $T_{\rm F}$ are the superconducting transition and Fermi temperatures, respectively. Throughout this paper, we set $k_{\rm B}=\hbar=1$ for simplicity. In this theory, the extrinsic contribution is of leading order, while the intrinsic mechanism is smaller by a factor of $(k_{\rm F}\xi_0)^{-1}$. 
Hence, understanding the extrinsic contribution is essential for correctly interpreting future experiments measuring the ATHE.
While the intrinsic ATHE has been widely investigated in the context of topological material science~\cite{tanaka2011symmetry,sato2016majorana,sato2017topological,nomura2012cross,qin2011energy,sumiyoshi_ATHE,shitade2014heat,goswami2015topological,nakai2017laughlin,nakai2020weyl,moriya}, studies of the extrinsic ATHE have been very limited.
Our work thus fills this gap for WSCs.

The rest of this paper is organized as follows.
We introduce the model of WSCs in Sec.~\ref{Sec2} and present the expression for the intrinsic contribution to the THC in Sec.~\ref{Sec3}.
In Sec.~\ref{Sec4}, we review the quasiclassical transport theory, and in Sec.~\ref{Sec5}, we calculate the nonequilibrium quasiclassical Green function (QGF) under temperature gradients to obtain the thermal conductivity.
The results of the low-temperature expansion analysis of the thermal conductivity are given in Sec.~\ref{Sec6}.
In Sec.~\ref{Sec7}, we associate the extrinsic ATHE with the formation of the impurity bands near the Fermi energy.
In Sec.~\ref{Sec8}, we discuss how line-nodal excitations influence the ATHE, focusing on the opposing effects that nodal excitations have on longitudinal and anomalous Hall transport.
Section~\ref{Sec9} provides a summary and conclusion.
Detailed derivations of several mathematically complex equations in the main text are given in the Appendices.

\begin{figure}[t]
\includegraphics[width=8.5cm]{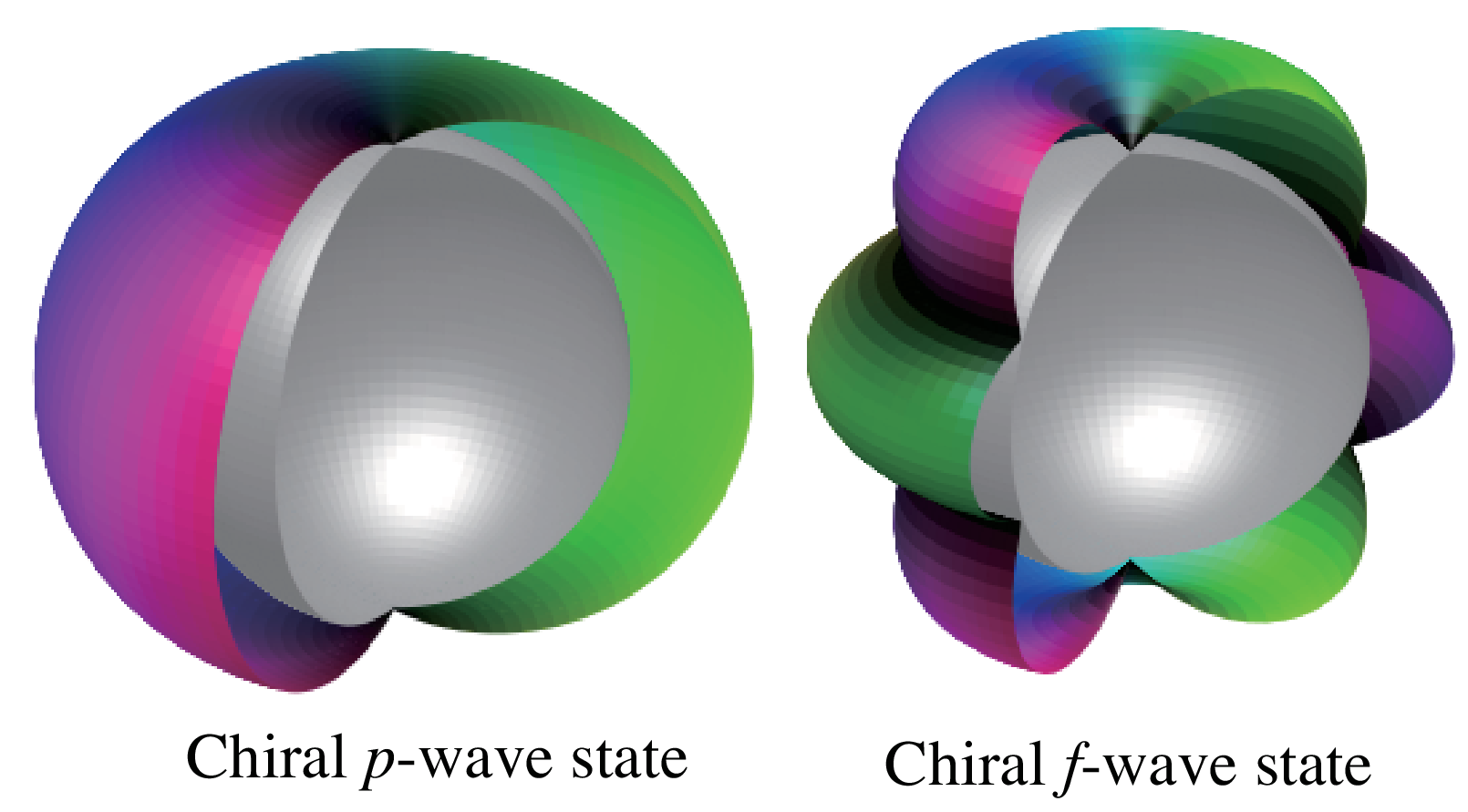}
\caption{
Superconducting gap structures in the $E_{1u}$ chiral superconducting states considered in the main text [see Eqs.~\eqref{sc_order} and \eqref{func_eta}].
The left and right panels illustrate the $E_{1u}$ chiral $p$-wave and $f$-wave states, respectively.
The color of the superconducting gap represents the phase of the complex order parameter.
The gray sphere indicates the underlying Fermi sphere.
} 
\label{fig1}
\end{figure}

\begin{table*}
\centering
\begin{tabular}{|c|c|c|c|c|}
    \hline
    Irreducible representation & Basis function & Winding number ($\nu$) & Parity &     \begin{tabular}{c}
    Extrinsic ATHE\\
    (the $\delta$-function type impurity potential)
    \end{tabular}\\
    \hline \hline
    $E_{1g}$ & $\hat{k}_z(\hat{k}_x\pm i\hat{k}_y)$ & $\pm1$ & + & $\times$~\cite{yilmaz2020spontaneous,ngampruetikorn2020impurity} \\
    \hline
    $E_{2g}$ & $(\hat{k}_x\pm i\hat{k}_y)^2$ & $\pm2$ & + & $\times$~\cite{yilmaz2020spontaneous,ngampruetikorn2020impurity} \\
    \hline
    $E_{1u}$ & $(\hat{k}_x\pm i\hat{k}_y) \hat{\bm {z}}$ & $\pm1$ & - & $\circ~$~\cite{yip2016low,ngampruetikorn2020impurity} \\
        \hline
    $E_{1u}$ & $(5\hat{k}_z^2-1)(\hat{k}_x\pm i\hat{k}_y)\hat{\bm {z}}$ & $\pm1$ & - & $\circ~$ \\
    \hline
    $E_{2u}$ & $\hat{k}_z(\hat{k}_x\pm i\hat{k}_y)^2 \hat{\bm {z}}$ & $\pm2$ & - &$\times$~\cite{ngampruetikorn2020impurity}\\
    \hline
\end{tabular}
\caption{The irreducible representations, the basis functions, the winding number ($\nu$), the parity, and the existence of the extrinsic ATHE due to the short-range impurity potential for the chiral superconducting order in the point group $D_{6h}$.}
\label{chiralorder_D6h}
\end{table*}

\section{Model}
\label{Sec2}
In this paper, we consider the chiral superconducting orders for the point group $D_{6h}$. 
Table.~\ref{chiralorder_D6h} summarizes the irreducible representations and their basis functions exhibiting the chiral superconducting order~\cite{RevModPhys.63.239}.
Among them, the skew scattering does not couple to the $E_{1g}$, $E_{2g}$ and $E_{2u}$ chiral superconducting order with the $\delta$-function type short-range impurity potential~\cite{yilmaz2020spontaneous,ngampruetikorn2020impurity}.
As described in Sec.~\ref{Sec1}, the $E_{2g}$ and $E_{2u}$ chiral superconducting orders have $|\nu|=2$, so that the skew scattering does not couple to these chiral orders.
The $E_{1g}$ chiral superconducting order has $|\nu|=1$, but the line-node at the equator ($k_z=0$) leads to a precise cancellation of the skew scattering contribution to the ATHE [see Sec.~V].
When impurities have finite size, $R\gtrsim 1/k_{\rm F}$, the Andreev scattering couples to these chiral superconducting orders and contributes to the ATHE.
Hence, the extrinsic mechanism is negligible in the $E_{1g}$, $E_{2g}$, and $E_{2u}$ chiral states when the impurity potential is short-range.

The $E_{1u}$ chiral superconducting order includes both $p$-wave and $f$-wave pairings.
As discussed below, these pairing states couple to the skew scattering, leading to the extrinsic ATHE.
In this paper, we examine the chiral $p$-wave and $f$-wave pairings in the $E_{1u}$ state on a spherical Fermi surface to evaluate the effects of impurity bands and nodal excitations on the skew scattering.
In these odd parity states, the $d$-vector is expressed by,
\begin{eqnarray}
    \label{sc_order}
    {\bm d}(\hat{\bm k})=\Delta (\hat{k}_x+i\hat{k}_y) \eta(\hat{k}_z)\hat{\bm z},
\end{eqnarray}
with the normalized crystal momentum $\hat{\bm k}={\bm k}/k_{\rm F}=(\hat{k}_x,\hat{k}_y,\hat{k}_z)$.
Here, $\hat{\bm z}$ represents the direction of the $d$-vector and $k_{\rm F}$ is the radius of the Fermi sphere.
The function $\eta(\hat{k}_z)$ has to be even $\eta(\hat{k}_z)=\eta(-\hat{k}_z)$, to satisfy the requirement ${\bm d}(\hat{\bm k})=-{\bm d}(-\hat{\bm k})$.
For the $E_{1u}$ chiral $p$-wave and $f$-wave pairing states, $\eta(\hat{k}_z)$ is given by,
\begin{equation}
\label{func_eta}
\eta(k_z)=
\begin{cases}
    1 & \text{chiral $p$-wave pairing},\\
    5\hat{k}_z^2-1 & \text{chiral $f$-wave pairing}.
\end{cases}
\end{equation}
These $E_{1u}$ chiral order parameters generate two Weyl points at the north and south poles on the Fermi sphere and realize WSCs with $\nu=1$.
As Fig.~\ref{fig1} shows, the $E_{1u}$ chiral $f$-wave pairing involves, in addition, the two horizontal line nodes at $\hat{k}_z=\pm 1/\sqrt{5}$.

These chiral order parameters are relevant to candidate materials of WSCs.
The chiral $p$-wave pairing, ${\bm d}(\hat{\bm k})=\Delta (\hat{k}_x\pm i\hat{k}_y) \hat{\bm z}$, is established in the ABM state of the superfluid $^3$He under an ambient pressure~\cite{leggett1975theoretical,sil14,mizushima2016symmetry,ikegamiscience,ikegamijpsj,shevtov}.
Although we consider the $D_{6h}$ point group, the calculated result with the chiral $p$-wave pairing is applicable to $^3$He because we do not consider the crystalline structure except for the form of the order parameter.
The $E_{1u}$ chiral $f$-wave pairing has been discussed as a candidate for the order parameter of UPt$_3$~\cite{tsutsumi2012spin,izawa2014pairing}.

\section{Intrinsic anomalous thermal Hall effect}
\label{Sec3}
In chiral superconductors, there are both intrinsic and extrinsic contributions to the ATHE.
One of our goals is to compare the two contributions to identify the dominant contribution.
In Sec.~\ref{Sec3}, we present the expression for the intrinsic contribution to the THC and show that its low-temperature behavior is completely determined by the configuration of Weyl points in momentum space.

For the model presented in Sec.~\ref{Sec2}, two Weyl points lie on the $k_z$ axis, separated by $\delta k_{\rm W}=2k_{\rm F}$ in momentum space.
Let us regard the WSCs as a family of two-dimensional superconductors labeled by $k_z$ in order to understand their topological nature.
Each subsystem labeled by $k_z$ is equivalent to a two-dimensional chiral superconductor and characterized by the $k_z$-resolved Chern number, 
\begin{eqnarray}
{\rm Ch}(k_z)=\int \frac{dk_xdk_y}{2\pi} \sum_{E_n(\bm k)<0} \mathcal{B}^n_{xy}(\bm k),
\end{eqnarray}
where $E_n(\bm k)$ is the quasiparticle energy and
\begin{eqnarray}
\mathcal{B}^n_{xy}(\bm k)=-2{\rm Im} \left\langle\frac{\partial u_n(\bm k)}{\partial k_x} \bigg| \frac{\partial u_n(\bm k)}{\partial k_y}\right\rangle,
\end{eqnarray} 
is the Berry curvature for the $n$-th band~\cite{RevModPhys.82.1959}.
When the effective mass is positive in the normal state, the $k_z$-resolved Chern number in the WSCs described by Eq.~\eqref{sc_order} is given by,
\begin{eqnarray}
\label{Chernnum}
{\rm Ch}(k_z)=2\Theta(k_{\rm F}^2-k_z^2),
\end{eqnarray}
where $\Theta(x)$ is the Heaviside step function~\cite{sato2016majorana,sato2017topological}.
The factor 2 in Eq.~\eqref{Chernnum} arises from the spin degrees of freedom.

The intrinsic contribution to the thermal Hall conductivity is given by the Berry curvature formula~\cite{qin2011energy,sumiyoshi_ATHE,shitade2014heat},
\begin{align}
\label{int_kxy_0}
\kappa_{xy}^{\rm int}=-\frac{1}{2T}\sum_n \int \frac{d \bm k}{(2\pi)^3} \int_{E_n(\bm k)}^{\infty} d\epsilon \epsilon^2
\mathcal{B}^n_{xy}(\bm k) \left(-\frac{\partial f}{\partial \epsilon}\right).
\end{align}
At low temperatures, $T\ll |\Delta|$, Eq.~\eqref{int_kxy_0} reduces to, 
\begin{eqnarray}
\label{int_kxy}
\kappa_{xy}^{\rm int}=\frac{\pi T}{6}\left( \frac{\delta k_{\rm W}}{2\pi}\right).
\end{eqnarray}
Equation~\eqref{int_kxy} clarifies that, at low temperatures, the intrinsic ATHE is completely determined by the configuration of Weyl points in momentum space~\cite{goswami2015topological,nakai2017laughlin,nakai2020weyl,moriya}. 
We note that this low-temperature formula does not depend on $\eta(\hat{k}_z)$, indicating that the intrinsic ATHE is insensitive to additional line-nodal excitations in WSCs at least to leading order in $T/T_c$.

The intrinsic ATHE is the first order in $(k_{\rm F}\xi_0)^{-1}$ and is not included in the quasiclassical transport (Eilenberger) theory~\cite{kobayashi2018negative}.
To illustrate this, we scale the thermal conductivity by $N(\epsilon_{\rm F})v_{\rm F}^2$ and, for our choice of $\delta k_{\rm W} = 2k_{\rm F}$, we find,
\begin{eqnarray}
\label{int_kxy_scaled}
    \frac{\kappa_{xy}^{\rm int}}{N(\epsilon_{\rm F})v_{\rm F}^2}=\frac{\pi}{12(k_{\rm F}\xi_0)}\left(\frac{T}{T_c}\right).
\end{eqnarray}
Equation~\eqref{int_kxy_scaled} shows that the intrinsic contribution appears only at the first order in $(k_{\rm F}\xi_0)^{-1}$, whereas the 
the standard quasiclassical transport theory keeps only the terms of leading (zeroth) order in $(k_{\rm F}\xi_0)^{-1}$ [see Eq.~\eqref{kappa_xy}].
Hence, the intrinsic contribution to the ATHE drops out in the quasiclassical Eilenberger formalism~\cite{Eilenberger, SERENE1983221}.

However, this hierarchy does not mean that the extrinsic contribution is always dominant. Note that, physically, the gapless surface Majorana modes are responsible for the intrinsic ATHE and always bring about a $T$-linear contribution in the THC at low temperature, regardless of the existence of nodal excitations in bulk~\cite{read2000paired,nomura2012cross,stone2012gravitational}. 
When the intrinsic contribution is derived from the bulk Hamiltonian using the Kubo theory, the effects of the surface Majorana mode are incorporated into the THC via a correction to the Kubo formula arising from the local equilibrium thermal magnetization current~\cite{qin2011energy,sumiyoshi_ATHE,shitade2014heat}.
In contrast, whether the extrinsic ATHE exhibits $T$-linear behavior in the low-temperature range, and, if it does, how large this contribution is, depends on the existence of nodal excitations and the details of the impurity scattering [see Ref.~\cite{yip2016low} and our analysis below]. Therefore, whether intrinsic or extrinsic contributions dominate at the low temperature depends on the specifics of the material.

\section{Quasiclassical transport theory}
\label{Sec4}
\subsection{Quasiclassical transport (Eilenberger) theory}
The quasiclassical transport (Eilenberger) theory describes superconductors in the limit $(k_{\rm F}\xi_0)^{-1}\ll 1$.	
In this limit, the normal state DOS can be taken to be energy-independent over the range where the Gor'kov Green function is peaked~\cite{Eilenberger}. 
Integrating the Gor'kov Green function over the band kinetic energy,
$\xi_{\bm k}={\bm k}^2/2m-\epsilon_{\rm F}$, where $\epsilon_{\rm F}$ is the Fermi energy, we define the quasiclassical Green function (QGF) as,
\begin{eqnarray}
\label{Keldysh_prop}
\check{g}(\epsilon,\bm k_{\rm F})&=&\int d\xi_{\bm k} \check{\tau}_z \check{G}(\epsilon,\bm k)\nonumber\\
&=&
\begin{pmatrix}
\underline{g}^{\rm R}(\epsilon,\bm k_{\rm F})&&\underline{g}^{\rm K}(\epsilon,\bm k_{\rm F})\\
0&&\underline{g}^{\rm A}(\epsilon,\bm k_{\rm F})
\end{pmatrix},
\end{eqnarray}
The QGF is defined at the Fermi surface and is the central object of the quasiclassical transport theory~\cite{Eilenberger, SERENE1983221}.
The superscript $\rm X=R, A, K$ in Eq.~\eqref{Keldysh_prop} represents the retarded, advanced, and Keldysh matrix elements, respectively. 
$\check{\tau}_\mu \; (\mu=x,y,z)$ are the Pauli matrices in the Nambu (particle-hole) space. 
Throughout this paper, we denote $\check{A}$ as a $8\times 8$ matrix in the Keldysh space and $\underline{A}$ as a $4\times 4$ Nambu matrix.
If a matrix $\underline{A}\; (\check{A})$ is defined in the Nambu (Keldysh) space, the corresponding matrix $\check{A}\; (\underline{A})$ in the Keldysh (Nambu) space is defined as $\check{A} = \underline{A}\otimes \openone$.

The QGF obeys the Eilenberger equation~\cite{Eilenberger,SERENE1983221},
\begin{eqnarray}
\label{Eilenberger}
\left[\epsilon \check{\tau}_z-\check{\Delta}-\check{\sigma}_{\rm imp},\check{g}\right]+i\bm v_{\rm F} \cdot {\bm \nabla} \check{g}=0,
\end{eqnarray}
and is supplemented by the normalization condition, 
\begin{eqnarray}
\check{g}^2=-\pi^2.    
\end{eqnarray}
This condition arises from the fact that the square of the QGF also satisfies the Eilenberger equation~\cite{Eilenberger, SERENE1983221}.
$\check{\Delta}$ is the superconducting order parameter matrix.
For spin-triplet superconductors with the $d$-vector, $\bm d(\bm k)$, this order parameter matrix is written as~\cite{SERENE1983221,RevModPhys.63.239},
\begin{subequations}
\begin{eqnarray}
\label{SCgap_mat}
\check{\Delta}&=&
\begin{pmatrix}
\underline{\Delta}&&\underline{0}\\
\underline{0}&&\underline{\Delta}
\end{pmatrix},\\
\underline{\Delta}&=&
\begin{pmatrix}
0&&i({\bm \sigma}\cdot {\bm d}(\bm k_{\rm F}))\sigma_y\\
i\sigma_y({\bm \sigma}\cdot {\bm d}^*(\bm k_{\rm F}))&&0
\end{pmatrix}.
\end{eqnarray}
where ${\bm \sigma}=(\sigma_x,\sigma_y,\sigma_z)$ is the vector of the Pauli matrices in the spin space.
$\check{\sigma}_{\rm imp}$ in Eq.~\eqref{Eilenberger} represents the impurity self-energy, which is expressed as,
\begin{eqnarray}
\check{\sigma}_{\rm imp}&=&
\begin{pmatrix}
\underline{\sigma}_{\rm imp}^{\rm R}&&\underline{\sigma}_{\rm imp}^{\rm K}\\
0&&\underline{\sigma}_{\rm imp}^{\rm A}
\end{pmatrix}.
\end{eqnarray}
\end{subequations}

In this paper, we consider nonmagnetic impurities with the short-range $\delta$-function potential, $V_{\rm imp}(\bm x)=\sum_{\bm R_{\rm imp}}V_{\rm imp}\delta(\bm x- \bm R_{\rm imp})$, where $\bm R_{\rm imp}$ is an impurity site and $V_{\rm imp}$ is the impurity potential strength.
The multiple scattering events are essential for the transverse transport and thus we compute the impurity self-energy with the self-consistent $T$-matrix approximation~\cite{yip2016low,yilmaz2020spontaneous,ngampruetikorn2020impurity}.
Assuming the random distribution of impurities and taking the impurity average, 
we obtain the self-consistent $T$-matrix equation~\cite{hirschfeld1988consequences,RevModPhys.78.373},
\begin{subequations}
\begin{eqnarray}
\label{tmateq1}
\check{\sigma}_{\rm imp}&=&n_{\rm imp}\check{t}_{\rm imp},\\
\label{tmat_eq2}
\check{t}_{\rm imp}&=&V_{\rm imp}
+N(\epsilon_{\rm F})V_{\rm imp}\braket{\check{g}}_{{\rm FS}}\check{t}_{\rm imp},
\end{eqnarray}
\end{subequations}
where $n_{\rm imp}$ is the impurity density and $N(\epsilon_{\rm F})$ is the DOS at the Fermi level in the normal state.
The bracket $\braket{\cdots}_{\rm FS}$ represents the normalized Fermi surface average, $\braket{1}_{\rm FS}=1$.
The impurity self-energy is independent of the Fermi momentum because we neglect the size of impurities.

We use the scattering rate $\Gamma_{\rm imp}=n_{\rm imp}/\pi N(\epsilon_{\rm F})$ and the scattering phase-shift $\cot \delta=-1/\pi N(\epsilon_{\rm F}) V_{\rm imp}$ in the normal state to parameterize disorder in our calculation.
Using these normal state quantities, we recast the self-consistent $T$-matrix equation as
\begin{eqnarray}
\label{tmat_swave}
\check{\sigma}_{\rm imp}=-\left[\cot \delta +\left\langle \frac{\check{g}}{\pi} \right\rangle_{\rm FS}\right]^{-1}\Gamma_{\rm imp}.
\end{eqnarray}
The limit $\delta \to 0\;(|\delta| \to \pi/2)$ corresponds to the Born (unitarity) limit, which represents the weak (strong) impurity potential limit, $V_{\rm imp}/\epsilon_{\rm F}\to 0$ ($|V_{\rm imp}/\epsilon_{\rm F}|\to \infty$).

\subsection{Response to Temperature Gradient}
The quasiclassical transport theory describes thermal responses~\cite{PhysRevB.53.15147, PhysRevB.75.224502}.
To include temperature gradients in the Eilenberger equation, we assume a local equilibrium, $T=T({\bm x})$, and expand the spatial gradient as $\bm \nabla=\bm \nabla T \frac{\partial}{\partial T}$.
We then obtain,
\begin{eqnarray}
\label{Keldysh_eq_w/_delT}
\left[ \epsilon\check{\tau}_z-\check{\Delta}-\check{\sigma}_{\rm imp},\check{g} \right]+(i{\bm v_{\rm F}}\cdot {\bm \nabla} T) \frac{\partial }{\partial T}\check{g}=0.
\end{eqnarray} 
The thermal current is calculated from the Keldysh QGF~as 
\begin{eqnarray}
    \label{thermal_current}
    {\bm J}_Q&=&N(\epsilon_{\rm F})\int \frac{d\epsilon}{4\pi i} \left\langle \frac{1}{4}{\rm Tr}\left[\epsilon {\bm v_{\rm F}} \underline{g}^{\rm K}\right]\right\rangle_{\rm FS}\,.
\end{eqnarray}
Once we compute the thermal current within 
the linear response theory, we define 
the thermal conductivity tensor,
\begin{eqnarray}
\label{kxyext}
J_{Qi}=\kappa^{\rm ext}_{ij}\left(-\partial_j T\right).
\end{eqnarray}
As discussed above in Sec.~\ref{Sec3}, in this framework, we only obtain the extrinsic contribution to the THC~\cite{kobayashi2018negative}.
Hence, we add the superscript ``{\rm ext}'' in Eq.~\eqref{kxyext} to represent the extrinsic (impurity-induced) contribution.

Once the quasiclassical limit is taken, the Berry curvature effects, including the anomalous velocity (effective Lorentz force in momentum space) and the side-jumps, drop out from the transport equation for the Green function~\cite{nagaosa2010anomalous,sinitsyn2007anomalous}.
As shown in Sec.~\ref{Sec3}, the Berry curvature effects appear from the first order in $(k_{\rm F}\xi_0)^{-1}$, and the gradient expansion to higher orders is required to account for these effects~\cite{kobayashi2018negative}.
Consequently, when we discuss the extrinsic ATHE using the Eilenberger equation, the ATHE originates from the skew scattering due to chiral Cooper pairs. 
When we compare the extrinsic ATHE with the intrinsic one, we refer to the low-temperature formula [Eq.~\eqref{int_kxy}].

\subsection{Quasiclassical Green function}
\label{Sec5}
Using the quasiclassical transport theory, we derive the QGF, which includes a linear response to the temperature gradient.
Everywhere below the equilibrium functions are denoted as $\check{x}_{\rm eq}\; (x=g,\; \Delta, \sigma_{\rm imp})$ and their linear deviation from the equilibrium are labeled as $\delta \check{x}\; (x=g,\; \Delta, \sigma_{\rm imp})$.

\subsubsection{Equilibrium quasiclassical Green function}
In global thermal equilibrium, the equilibrium QGF, $\check{g}_{\rm eq}$, is the solution of Eq.~\eqref{Eilenberger} without the gradient term. 
Because the Fermi surface average for the order parameter matrix vanishes in $p$- and $f$-wave superconductors, $\braket{\underline{\Delta}(\bm k_{\rm F})}_{\rm FS}=0$, the equilibrium impurity self-energy is diagonal in the Nambu space.
Then, the Eilenberger equation can be easily solved using the normalization condition to give, 
\begin{eqnarray}
\label{g_eq_RA}
\underline{g}_{\rm eq}^{\rm R, A}&=&-\pi \frac{\underline{M}^{\rm R,A}}{D^{\rm R,A}},\\
\label{g_eq_K}
\underline{g}_{\rm eq}^{\rm K}&=&\left( \underline{g}_{\rm eq}^{\rm R}-\underline{g}_{\rm eq}^{\rm A} \right)\tanh \left(\frac{\epsilon}{2T}\right),
\end{eqnarray}
where $\underline{M}^{\rm R,A}=\tilde{\epsilon}^{\rm R,A}\underline{\tau}_z-\underline{\Delta}_{\rm eq}$, $D^{\rm R,A}=\sqrt{|{\bm d}({\bm k}_{\rm F})|^2-\tilde{\epsilon}^{{\rm R,A}\;2}}$ and $\tilde{\epsilon}^{\rm R,A}=\epsilon-\frac{1}{4}{\rm Tr}(\underline{\tau}_z\underline{\sigma}_{\rm imp,eq}^{\rm R,A})$.

\subsubsection{Nonequilibrium quasiclassical Green function}
The nonequilibrium QGF, $\delta \check{g}$, which describes the linear response to the temperature gradient, obeys the Eilenberger equation, 
\begin{eqnarray}
    \label{noneq_Eilenberger}
    &\left[\epsilon \check{\tau}_z-\check{\Delta}_{\rm eq}-\check{\sigma}_{\rm imp, eq},\delta\check{g}\right]
    -\left[\delta\check{\sigma}_{\rm imp},\check{g}_{\rm eq}\right]\nonumber\\
    &+i\bm v_{\rm F} \cdot {\bm \nabla} T \frac{\partial}{\partial T} \check{g}_{\rm eq}=0.
\end{eqnarray}
We can straightforwardly solve Eq.~\eqref{noneq_Eilenberger} for the  retarded and the advanced QGFs. Using the normalization condition $\{\underline{g}^{\rm R,A}_{\rm eq},  \delta \underline{g}^{\rm R,A}\}=0$, we obtain,
\begin{eqnarray}
\label{del g^X}
\delta \underline{g}^{\rm R,A}=\frac{\underline{g}^{\rm R,A}_{\rm eq}}{2\pi D^{\rm R,A}}\left( \left[\delta \underline{\sigma}^{\rm R,A}_{\rm imp},\underline{g}^{\rm R,A}_{\rm eq}  \right]-(i{\bm v_{\rm F}}\cdot {\bm \nabla} T) \frac{\partial}{\partial T} \underline{g}^{\rm R,A}_{\rm eq} \right).\nonumber\\
\end{eqnarray}
The equilibrium retarded and advanced QGFs depend on the temperature only through the gap function, which makes the second term in Eq.~\eqref{del g^X} negligible at low temperatures. That term is traceless, as can be easily verified using $(\underline{g}^{\rm R, A}_{\rm eq})^2=-\pi^2$~\cite{yip2016low, PhysRevB.53.15147},
\begin{eqnarray}
{\rm Tr}\left[\underline{g}^{\rm R,A}_{\rm eq} \frac{\partial}{\partial T} \underline{g}^{\rm R,A}_{\rm eq}\right]&=& 2{\rm Tr}\left[\frac{\partial}{\partial T} 
(\underline{g}^{{\rm R,A}}_{\rm eq})^2\right]=0.
\end{eqnarray} 

For the Keldysh component, it is convenient to define the anomalous Keldysh QGF, $\delta \underline{g}^{a}$, and the anomalous Keldysh impurity self-energy, $\delta \underline{\sigma}_{\rm imp}^{a}$,
\begin{eqnarray}
\label{ga}
\delta \underline{g}^{\rm K}&=&\left(\delta \underline{g}^{\rm R}-\delta \underline{g}^{\rm A}\right)\tanh \left(\frac{\epsilon}{2T}\right)+\delta \underline{g}^{a},\\
\delta \underline{\sigma}_{\rm imp}^{K}&=&\left(\delta \underline{\sigma}_{\rm imp}^{\rm R}-\delta \underline{\sigma}_{\rm imp}^{\rm A}\right)\tanh \left(\frac{\epsilon}{2T}\right)+\delta \underline{\sigma}_{\rm imp}^{\rm a}.
\end{eqnarray}
The first term in Eq.~\eqref{ga} describes the change in the spectral function, which is proportional to $\underline{g}^{\rm R}- \underline{g}^{\rm A}$ while maintaining the distribution in equilibrium. This term does not contribute to the thermal conductivity~\cite{PhysRevB.53.15147}.
The second term in Eq.~\eqref{ga} accounts for the change in the distribution function from equilibrium and is essential for evaluating thermal transport properties.

This separation of the nonequilibrium Keldysh QGF simplifies the transport equation and allows us to solve the Eilenberger equation for the Keldysh component.
Using the constraint $\underline{g}^{{\rm R}}_{\rm eq} \delta  \underline{g}^{{a}} +\delta  \underline{g}^{{a}}\underline{g}^{{\rm A}}_{\rm eq}=0$ from the normalization condition, we obtain,
\begin{subequations}
\begin{eqnarray}
\label{gEli}
\delta \underline{g}^{a}&=&\delta \underline{g}^{a}_{\rm ns}+\delta \underline{g}^{a}_{\rm vc},\\
\label{gEli_ns}
\delta \underline{g}^{a}_{\rm ns}&=&\underline{N}^{\rm R}_{\rm eq}\left( \underline{g}_{\rm eq}^{\rm R}-\underline{g}_{\rm eq}^{\rm A}\right)\left(-\frac{i\left({\epsilon \bm v_{\rm F}}\cdot {\bm \nabla} T\right)}{2T^2\cosh^2\left(\frac{\epsilon}{2T}\right)}\right),\\
\label{gEli_vc}
\delta \underline{g}^{a}_{\rm vc}&=&
\underline{N}^{\rm R}_{\rm eq}\left(\underline{g}_{\rm eq}^{\rm R} \delta \underline{\sigma}_{\rm imp}^{a}- \delta \underline{\sigma}_{\rm imp}^{a}\underline{g}_{\rm eq}^{\rm A} \right),
\end{eqnarray}
\end{subequations}
where we defined the retarded function as,
\begin{eqnarray}
\label{NR}
\underline{N}^{\rm R}_{\rm eq}=\frac{\left(D^{\rm R}+D^{\rm A}\right)\left(-\frac{\underline{g}^{\rm R}_{\rm eq}}{\pi}\right)+\sigma_{\rm imp,eq0}^{\rm R}-\sigma_{\rm imp,eq0}^{\rm A} }{\left(D^{\rm R}+D^{\rm A} \right)^2+\left( \sigma_{\rm imp,eq0}^{\rm R}-\sigma_{\rm imp,eq0}^{\rm A} \right)^2},
\end{eqnarray}
with the trace of the equilibrium self-energy $\sigma_{\rm imp,eq0}^{\rm R,A}={\rm Tr}(\sigma_{\rm imp,eq}^{\rm R,A})$. Note that
$\delta \underline{g}^{a}_{\rm vc}$ involves $\delta \underline{\sigma}_{\rm imp}^{a}$, which includes the vertex correction in the diagrammatic calculations.
We thus refer to $\delta \underline{g}^{a}_{\rm ns}$ as a non-selfconsistent contribution and $\delta \underline{g}^{a}_{\rm vc}$ as a vertex correction contribution, respectively.

\section{Thermal conductivity at low temperature}
\label{Sec6}
As shown in Eq.~(23), the anomalous Keldysh QGF is proportional to the derivative of the Fermi distribution function, $\partial_\epsilon f(\epsilon)\propto 1/\cosh^2(\epsilon/2T)$.
This factor peaks at $\epsilon=0$ and introduces the frequency cut-off $\epsilon_{\rm cut} \sim T$.
This cut-off frequency becomes small at low temperatures, justifying
the expansion of the QGF in $\epsilon$~\cite{yip2016low}.
We perform this low-temperature expansion to obtain the low-temperature formula for the extrinsic ATHE.
    
\begin{figure}[t]
    \includegraphics[width=8.7cm]{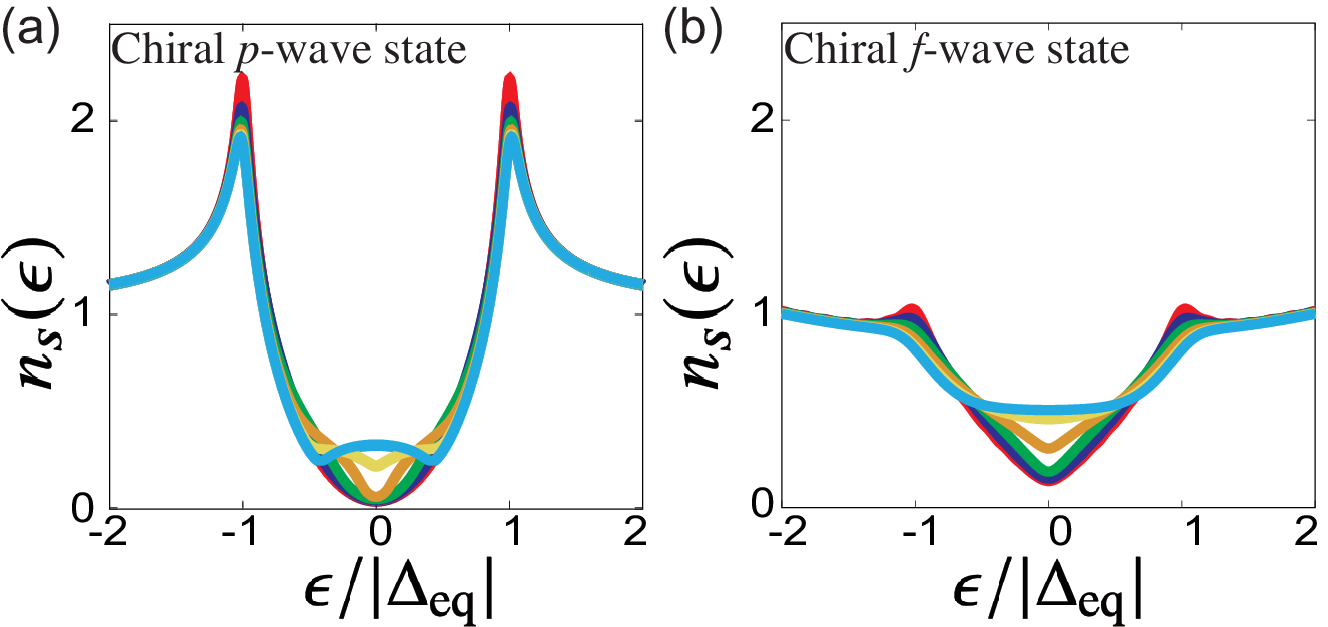}
    \caption{Quasiparticle DOS in (a) the chiral $p$-wave and (b) the chiral $f$-wave states for the several values of the scattering phase-shift.
    In this calculation, we set the scattering rate $\Gamma_{\rm imp}=0.04\pi T_{c,\rm clean}$ ($T_{c,\rm clean}$ is a critical temperature in clean systems) and the scattering phase-shift $\delta=\frac{\pi}{12}$(red curves), $\frac{\pi}{6}$(blue curves), $\frac{\pi}{4}$(green curves), $\frac{\pi}{3}$(orange curves), $\frac{5\pi}{12}$(yellow curves), $\frac{\pi}{2}$(sky blue curves).} 
\label{fig2}
\end{figure}
        
Following the procedure outlined in Ref.~\cite{yip2016low}, we obtain the low-temperature formula for the thermal conductivity as~\cite{appendix},
\begin{align}
\frac{\kappa^{\rm ext}_{yy}}{N(\epsilon_{\rm F})v_{\rm F}^2 } 
=&\frac{\pi^2 T}{6}\gamma^2 \braket{\alpha_0(\hat{k}_z)}_{{\rm FS}}\nonumber\\
& +\frac{\pi^2 \Gamma_{\rm imp}\gamma^2|\Delta_{\rm eq}|^2T}{3}  Y\braket{\alpha_1(\hat{k}_z)}_{{\rm FS}}^2 \nn \\
& + \mathcal{O}(T^2,\Gamma_{\rm imp}^4),
\label{kappa_yy} \\
\frac{\kappa^{\rm ext}_{xy}}{N(\epsilon_{\rm F})v_{\rm F}^2}
=&-\frac{\pi^2 \Gamma_{\rm imp}\gamma^2|\Delta_{\rm eq}|^2T}{3} X \braket{\alpha_1(\hat{k}_z)}_{{\rm FS}}^2 \nn \\
& +\mathcal{O}(T^2,\Gamma_{\rm imp}^4),
\label{kappa_xy}
\end{align}
with $i\gamma \equiv -\frac{1}{4}{\rm Tr}\left[\underline{\tau}_z\underline{\sigma}_{\rm imp,eq}^{\rm R}(\epsilon=0)\right]$.
In Appendix~\ref{sec:app_LTE}, we derive the complete expression for the thermal conductivity at the low temperature, including higher-order terms in the scattering rate, $\Gamma_{\rm imp}$. 
In Eqs.~\eqref{kappa_yy} and \eqref{kappa_xy}, the dimensionless factors $X$ and $Y$ are defined as,
\begin{subequations}
\begin{eqnarray}
X&=& \frac{n_s(0) \cot \delta  }{2\left(\cot^2 \delta +n_s^2(0)\right)^2 },\\
Y&=& \frac{\cot^2 \delta -n_s^2(0) }{4\left(\cot^2 \delta +n_s^2(0)\right)^2 },
\end{eqnarray}
\end{subequations}
with $n_s(\epsilon)=\frac{N_s(\epsilon)}{N(\epsilon_{\rm F})}$.
$N_s(\epsilon)$ is the quasiparticle DOS,
\begin{eqnarray}
N_s(\epsilon)=N(\epsilon_{\rm F})\left\langle -\frac{1}{4}{\rm Tr}{\rm Im}\left(\frac{\underline{g}^{\rm R}_{\rm eq}(\epsilon)}{\pi}\right)\right\rangle_{\rm FS}\,.
\end{eqnarray}
Additionally, the Fermi surface functions $\alpha_n(\hat{k}_z)$ $(n\in \mathbb{Z})$ are defined as
\begin{eqnarray}
\label{alpha}
\alpha_n(\hat{k}_z)&=&\frac{ \hat{k}_{\perp}^2 \eta^n(\hat{k}_z)}{\left(\gamma^2 + |\Delta_{\rm eq} \hat{k}_{\perp} \eta(\hat{k}_z)|^2\right)^{\frac{3}{2}}},
\end{eqnarray}
with $\hat{k}_{\perp}^2=\hat{k}_{x}^2+\hat{k}_{y}^2=1-\hat{k}_{z}^2$.
$\eta(\hat{k}_z)$ is defined in Sec.~\ref{Sec2} and describes the additional nodal and near-nodal structures.

\begin{figure}[t]
    \includegraphics[width=8.7cm]{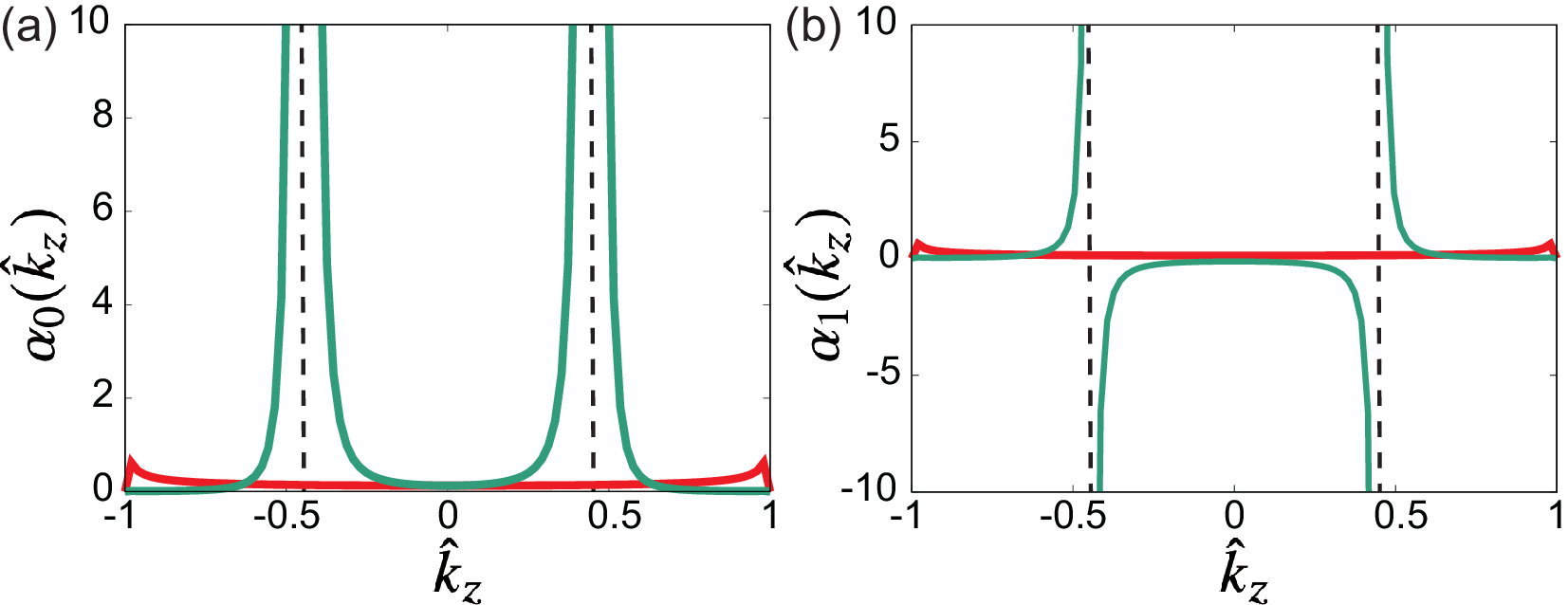}
    \caption{Fermi surface functions (a) $\alpha_0(\hat{k}_z)$ and (b) $\alpha_1(\hat{k}_z)$ in the chiral $p$-wave and the $f$-wave states. 
    In all panels, the red and green curves represent the calculated results for the chiral $p$-wave and $f$-wave states, respectively.
    The positions of the line-nodes in the chiral $f$-wave state ($\hat{k}_z=\pm 1/\sqrt{5}$) are indicated by the vertical dashed lines.
    In these calculations, we set the scattering rate $\Gamma_{\rm imp}=0.04\pi T_{c,\rm clean}$ and the phase-shift $\delta=\frac{\pi}{6}$.}
    \label{fig4}
\end{figure}

The factors $X$ and $Y$ stem from the vertex correction contributions to the nonequilibrium Keldysh QGF.
These factors are closely related to impurity-bound states in superconductors, and we briefly review 
the common physics of these states before discussing the thermal transport properties~\cite{PhysRevLett.128.097001,matsushita2024spin}.
In unconventional superconductors with a momentum-dependent gap function, multiple Andreev scattering at impurity sites creates quasiparticle-bound states.
In our formulation, such impurity effects are included in the impurity self-energy under the self-consistent $T$-matrix approximation.
The impurity self-energy introduces new poles in the equilibrium QGF.
These new poles occur at both positive (electron-like) and negative (hole-like) energies due to the particle-hole symmetric spectrum. 
The impurity level is positioned at $\epsilon=0$ in the unitarity limit ($|\delta|=\pi/2$), and moves to a finite energy when the phase-shift deviates from this limit.
For the weak impurity potential ($|\delta|\ll \pi/2$), the spectral weight for the bound states merges with the coherence peaks in the DOS at $\epsilon = \pm |\Delta_{\rm eq}|$.
For a finite impurity density, the impurity-bound states broaden into impurity bands, centered at the energy of single impurity resonant states, with a finite bandwidth determined by the details of impurity scattering.

Figure~\ref{fig2} shows the DOS for the several values of $\delta$, illustrating the phase-shift dependence explained above.
When the impurity potential is weak (such as the red, green and blue curves in Fig.~\ref{fig2} (a-b)), the low-energy DOS ($|\epsilon|= |\Delta_{\rm eq}|$) varies with the energy as $N_s(\epsilon)\propto \epsilon^2 $ for the chiral $p$ wave state, and $N_s(\epsilon)\propto |\epsilon| $ for the chiral $f$ wave state, reflecting the dimensionality of the nodal regions in the gap structure~\cite{RevModPhys.63.239}.
As $\delta$ increases, the spectral weight in the DOS is partially transferred from the coherence peaks at $|\epsilon|= |\Delta_{\rm eq}|$ into the mid-gap region $|\epsilon|< |\Delta_{\rm eq}|$.
As the scattering phase-shift approaches the unitarity limit, even in the dilute impurity limit that we consider, the transferred spectral weight shifts close to the Fermi energy, with the tails of the impurity bands touching $\epsilon=0$ [see the orange, yellow, and sky blue curves in Fig.~\ref{fig2} (a-b)].

\begin{figure}[t]
    \includegraphics[width=8.7cm]{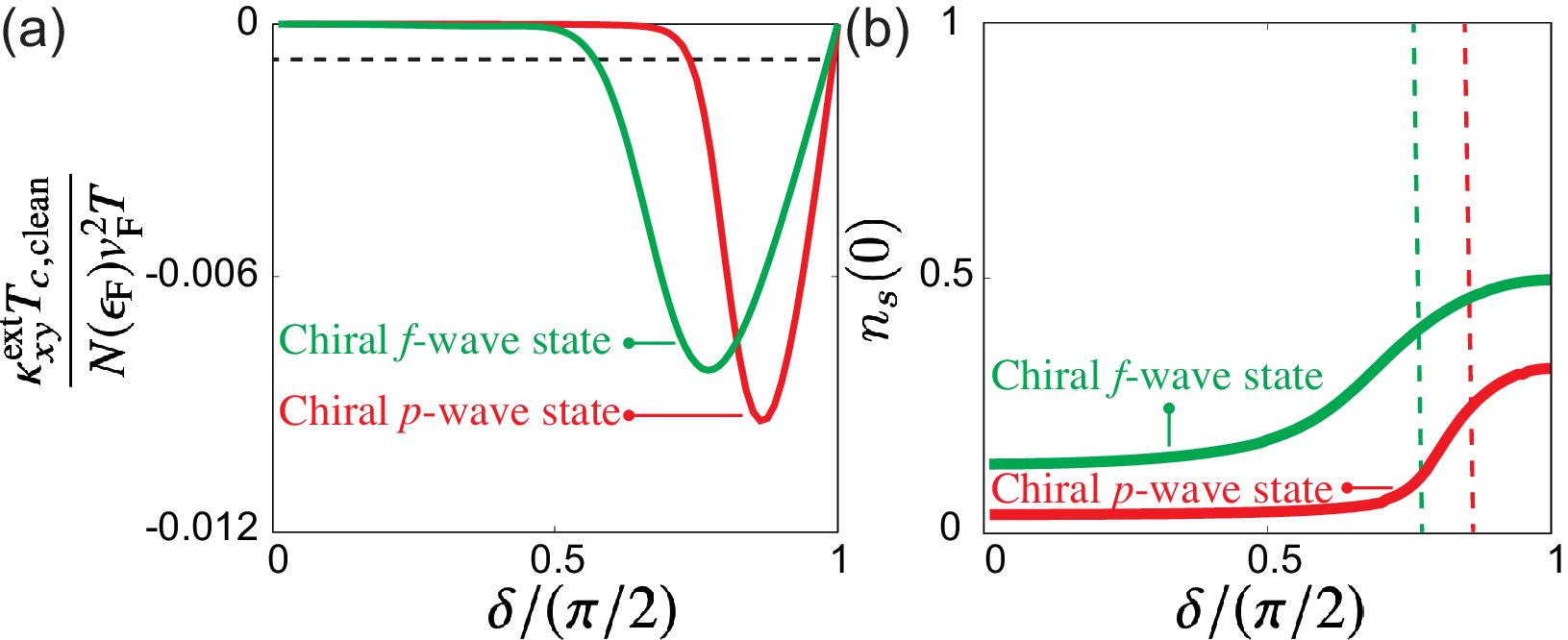}
    \caption{Scattering phase-shift dependence of (a) the impurity contribution to the THC, and (b) the zero-energy DOS in the $E_{1u}$ chiral $p$-wave and $f$-wave states at the low temperature.
    In all panels, The red and green curves represent the calculated results for the chiral $p$-wave and $f$-wave states. 
    In panel (a), the dotted line plots $-\kappa_{xy}^{\rm int}T_{c,{\rm clean}}/(N(\epsilon_{\rm F})v_{\rm F}^2T)$ with $(k_{\rm F}\xi_0)^{-1}=0.01$, where the minus sign is due to compare the relative magnitude with $\kappa_{xy}^{\rm ext}$.
    In panel (b), the dashed vertical lines indicate the peak positions of the THC as a function of the scattering phase-shift for the chiral $p$-wave state (red line) and the chiral $f$-wave state (green line).
    In these calculations, we set the scattering rate $\Gamma_{\rm imp}=0.04\pi T_{c,{\rm clean}}$ and the temperature $T=0.02T_{c,{\rm clean}}$.
    }
    \label{fig3}
\end{figure}

\begin{figure*}[t]
    \includegraphics[width=18cm]{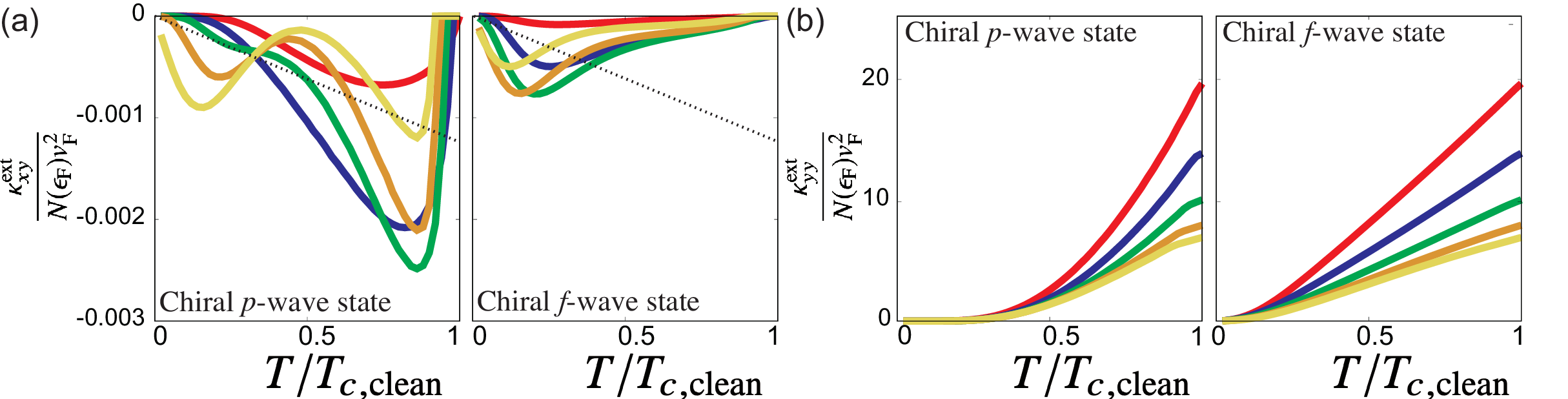}
    \caption{Temperature dependence of (a) the impurity contribution to the THC and (b) the longitudinal thermal conductivity in the $E_{1u}$ chiral $p$-wave and $f$-wave states.
    In panel (a), the dotted lines plot $-\kappa_{xy}^{\rm int}/(N(\epsilon_{\rm F})v_{\rm F}^2)$ using the low-temperature formula with $(k_{\rm F}\xi_0)^{-1}=0.01$, where the minus sign is due to comparing the relative magnitude with $\kappa_{xy}^{\rm ext}$.
    In all panels, we set the scattering rate $\Gamma_{\rm imp}=0.04\pi T_{c,{\rm clean}}$ and the scattering phase-shift $\delta=\frac{\pi}{12}$ (red curves), $\frac{\pi}{6}$ (blue curves), $\frac{\pi}{4}$ (green curves), $\frac{\pi}{3}$ (orange curves), $\frac{5\pi}{12}$ (yellow curves). 
    }
        \label{fig5}
\end{figure*}

We now focus on the thermal transport coefficients at low temperatures, including their sign and magnitude.
The impurity contribution to the THC is proportional to the factor $X$, which has a positive (negative) value when the impurity potential is attractive (repulsive) for electrons.
Referring to the low-temperature formula for the intrinsic contribution [Eq.~\eqref{int_kxy}], we find that the intrinsic and impurity contributions to the THC have opposite signs for the attractive impurity potential ($0<\delta<\pi/2$)~\cite{yip2016low}.

The first term in Eq.~\eqref{kappa_yy} is proportional to $\braket{\alpha_0(\hat{k}_z)}_{{\rm FS}}$.
As shown in Fig.~\ref{fig4} (a), $\alpha_0(\hat{k}_z)$ is positive and sharply peaks at the positions of the line-nodes on the Fermi surface, making its Fermi surface average large. 
Hence, line-nodal excitations always enhance the longitudinal thermal transport [see also Fig.~\ref{fig5} (b)].
In contrast, whether line-nodal excitations enhance the extrinsic ATHE depends on the details of the order parameters [see also Sec.~\ref{Sec8}].
The impurity contribution to the THC is proportional to the square of the Fermi surface average of $\alpha_1(\hat{k}_z)$. $\alpha_1(\hat{k}_z)$ changes sign across line-nodes due to $\eta(\hat{k}_z)$ in the numerator, which reduces its average over the Fermi surface.
Figure~\ref{fig4} (b) clearly shows that in the chiral $f$-wave state, the two horizontal line-nodes involve such sign changes in $\alpha_1(\hat{k}_z)$.
As we show in Sec.~\ref{Sec8}, this sign change at line nodes significantly affects the extrinsic ATHE.

Because only unpaired quasiparticles carry entropy, a finite residual DOS is a necessary, but not sufficient, condition to obtain a finite impurity contribution to the THC [see Eq.~\eqref{kappa_xy}]. 
As shown in Fig.~\ref{fig3}~(a), the impurity contribution to the THC vanishes both in the Born $(\delta \to 0)$ and unitarity $(|\delta| \to \pi/2)$ limits and exhibits a peak at the intermediate the scattering phase-shift.
In the case of $\Gamma_{\rm imp}=0.04\pi T_{c,{\rm clean}}$, the corresponding values of the phase-shift are $\delta \simeq 0.43\pi$ for the chiral $p$-wave state and $\delta \simeq 0.39\pi$ for the chiral $f$-wave state.
The key observation is that as other transverse transport coefficients, the extrinsic ATHE is enhanced by the effective DOS anisotropy at the Fermi surface, i.e., the finite slope of the DOS as a function of energy near $\epsilon=0$~\cite{PhysRevLett.128.097001,matsushita2024spin}. 
This slope vanishes in both Born and unitarity limits as shown in Fig.~\ref{fig2}, but $(\partial N_s/\partial \epsilon)_{\epsilon=0} \neq 0$ is realized in the scattering phase-shifts between these limits. 
In Sec.~\ref{Sec7}, we clarify the relation between the extrinsic ATHE and the emergence of the impurity bands.

Figure~\ref{fig3} (a) shows that even at low temperatures, the impurity contribution to the THC becomes larger than the intrinsic contribution when the scattering phase-shift is within a certain range.
For $\Gamma_{\rm imp}=0.04\pi T_{c,\rm clean}$, these ranges are $0.382\pi \lesssim \delta \lesssim 0.497\pi$ for the chiral $p$-wave state, and $0.284\pi \lesssim \delta \lesssim 0.490\pi$ for the chiral $f$-wave state.
This result suggests an underlying competition between the intrinsic and extrinsic ATHE, even at low temperatures.

\section{Impurity bound states and anomalous thermal Hall effect}
\label{Sec7}
We now explicitly associate the extrinsic ATHE to the formation of impurity bands in unconventional superconductors.
From the $T$-matrix equation in equilibrium and the residual DOS, we find the relationship between the extrinsic ATHE and the residual DOS,
\begin{eqnarray}
\label{ATHE_DOS}
\frac{\kappa^{\rm ext}_{xy}}{N(\epsilon_{\rm F})v_{\rm F}^2 } &=&-\frac{\pi^2\Gamma_{\rm imp} \gamma^2T}{12}\frac{\braket{\alpha_1(\hat{k}_z)}_{{\rm FS}}^2}{\braket{\alpha_2(\hat{k}_z)}_{\rm FS}} \frac{\partial n_s(0)}{\partial \delta}\sin^2 \delta \nonumber\\
&+&\mathcal{O}(T^2,\Gamma_{\rm imp}^4).
\end{eqnarray} 
Equation~\eqref{ATHE_DOS} is derived in Appendix~\ref{sec:app32}.
As the scattering phase-shift increases, the zero-energy DOS begins to grow when the tails in the DOS from the electron-like and hole-like impurity bands touch the Fermi level [see the orange curves Fig.~\ref{fig2} (a-b)].
This increase in $n_s(0)$ with $\delta$ eventually saturates in the unitarity limit [see Fig.~\ref{fig3} (b)].
The factor, $\partial n_s(0)/\partial \delta$, captures this variation in the residual DOS due to the change in the impurity levels and links the ATHE with the formation of the impurity bands.

\section{Line-nodes and Anomalous Thermal Hall Effect}
\label{Sec8}
\subsection{Chiral $p$-wave state v.s. chiral $f$-wave state}
We are now in the position to compare the full numerical results for the thermal conductivity in the $E_{1u}$ chiral $p$-wave and $f$-wave states, evaluating the influence of additional line nodes on the extrinsic ATHE.

Figure~\ref{fig5} (a) shows the temperature dependence of the impurity contribution to the THC.
As predicted by the low-temperature expansion analysis, the maximum amplitude of $\kappa_{xy}^{\rm ext}$ is small when the impurity potential strength is close to the Born or unitarity limit [see the red and yellow curves in Fig.~\ref{fig5} (a)], and becomes large at intermediate phase-shifts between these limits.
The relative magnitude of the intrinsic and extrinsic ATHE is consistent with the result of the low-temperature expansion analysis, see Figs.~\ref{fig3} (a) and~\ref{fig5} [see the green, blue, and orange curves in Fig.~\ref{fig5} (a)].
If we assume that the intrinsic (topological) contribution to the THC discussed in Sec.~\ref{Sec3} maintains its $T$-linear growth, the intrinsic mechanism may dominate the extrinsic (impurity-induced) contribution at higher temperatures.
However, as the temperature increases, corrections to the result in Eq.~\eqref{int_kxy} are generally expected.

\begin{figure}[b]
    \includegraphics[width=7cm]{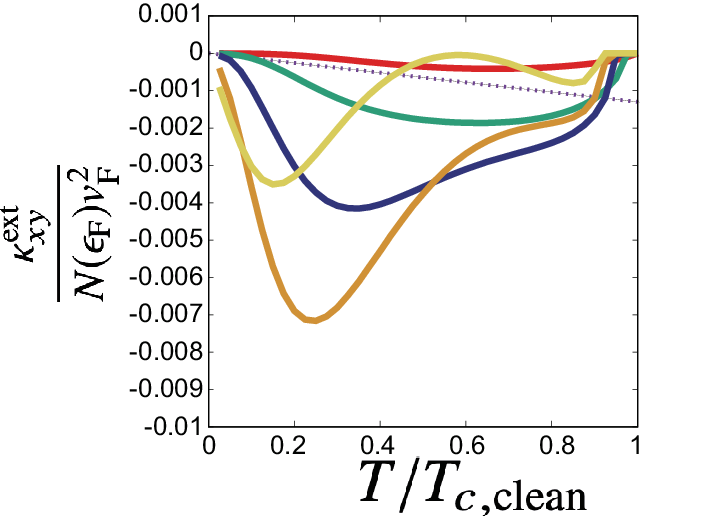}
    \caption{ 
    Temperature dependence of the impurity contribution to the THC in the chiral $f$-wave states described by Eq.~\eqref{abs_chiral}.
    The dotted lines indicate $-\kappa_{xy}^{\rm int}$ using the low-temperature formula with $(k_{\rm F}\xi_0)^{-1}=0.01$, while the minus sign enables direct comparison with the extrinsic term, $\kappa_{xy}^{\rm ext}$.
    We set the scattering rate $\Gamma_{\rm imp}=0.04\pi T_{c,{\rm clean}}$ and the scattering phase-shift $\delta=\frac{\pi}{12}$ (red lines), $\frac{\pi}{6}$ (blue lines), $\frac{\pi}{4}$ (green lines), $\frac{\pi}{3}$ (orange lines), $\frac{5\pi}{12}$ (yellow lines).} 
    \label{test}
\end{figure}

Notably, the maximum amplitude of $\kappa_{xy}^{\rm ext}$ in the $E_{1u}$ chiral $f$-wave state with two horizontal line-nodes is smaller than that of the $E_{1u}$ chiral $p$-wave state. 
This is in sharp contrast with the longitudinal thermal transport, which is enhanced by the enlarged phase space available near nodal lines [see Fig.~\ref{fig5} (b)]. 
This result implies that the extrinsic ATHE is weakened by additional line nodes in the chiral $f$-wave state.
This observation is consistent with the low-temperature expansion analysis discussed in Sec.~\ref{Sec6}. 
The function $\alpha_1(\hat{k}_z)$ is proportional to $\eta(\hat{k}_z)$ and changes sign across the additional line-nodes, as illustrated in Fig.~\ref{fig4} (b). 
This sign change reduces the Fermi surface average of $\alpha_1(\hat{k}_z)$ and suppresses the extrinsic anomalous thermal Hall response.

Let us demonstrate that the sign changes at the line nodes suppress the extrinsic ATHE in the $E_{1u}$ chiral $f$-wave state. 
To this end, we consider the chiral $f$-wave order parameter on the Fermi sphere,
\begin{eqnarray}
\label{abs_chiral}
    {\bm d}(\bm k_{\rm F})=\Delta(\hat{k}_x+i\hat{k}_y) |5\hat{k}_z^2-1|\hat{\bm z}.
\end{eqnarray}
This chiral state has the same quasiparticle population as the $E_{1u}$ chiral $f$-wave state because this state has two-horizontal line-nodes at $\hat{k}_z=\pm 1/\sqrt{5}$ and Weyl points at the north and south poles on the Fermi sphere.
However, the sign changes at the line nodes are eliminated by taking the absolute value in Eq.~\eqref{abs_chiral}.
Figure~\ref{test} shows that the chiral state described by Eq.~\eqref{abs_chiral} exhibits a larger ATHE signal than the $E_{1u}$ chiral $p$-wave and $f$-wave states [see Fig.~\ref{fig5} (a)].
This is because $\langle \alpha_1(\hat{k}_z)\rangle_{\rm FS}$ becomes large by taking the absolute value in Eq.~\eqref{abs_chiral}.

We also note that the $E_{1g}$ chiral $d$-wave state with $\Delta(\bm k)\propto k_z(k_x\pm ik_y)$ has a line node at the equator, so that the extrinsic ATHE is suppressed. 
In this case, the extrinsic contributions from the northern and southern hemispheres precisely cancel each other, and the intrinsic contribution dominates the thermal Hall response.

\subsection{Coexistence of the chiral $p$-wave and $f$-wave pairings}
\begin{figure*}[t]
    \includegraphics[width=18cm]{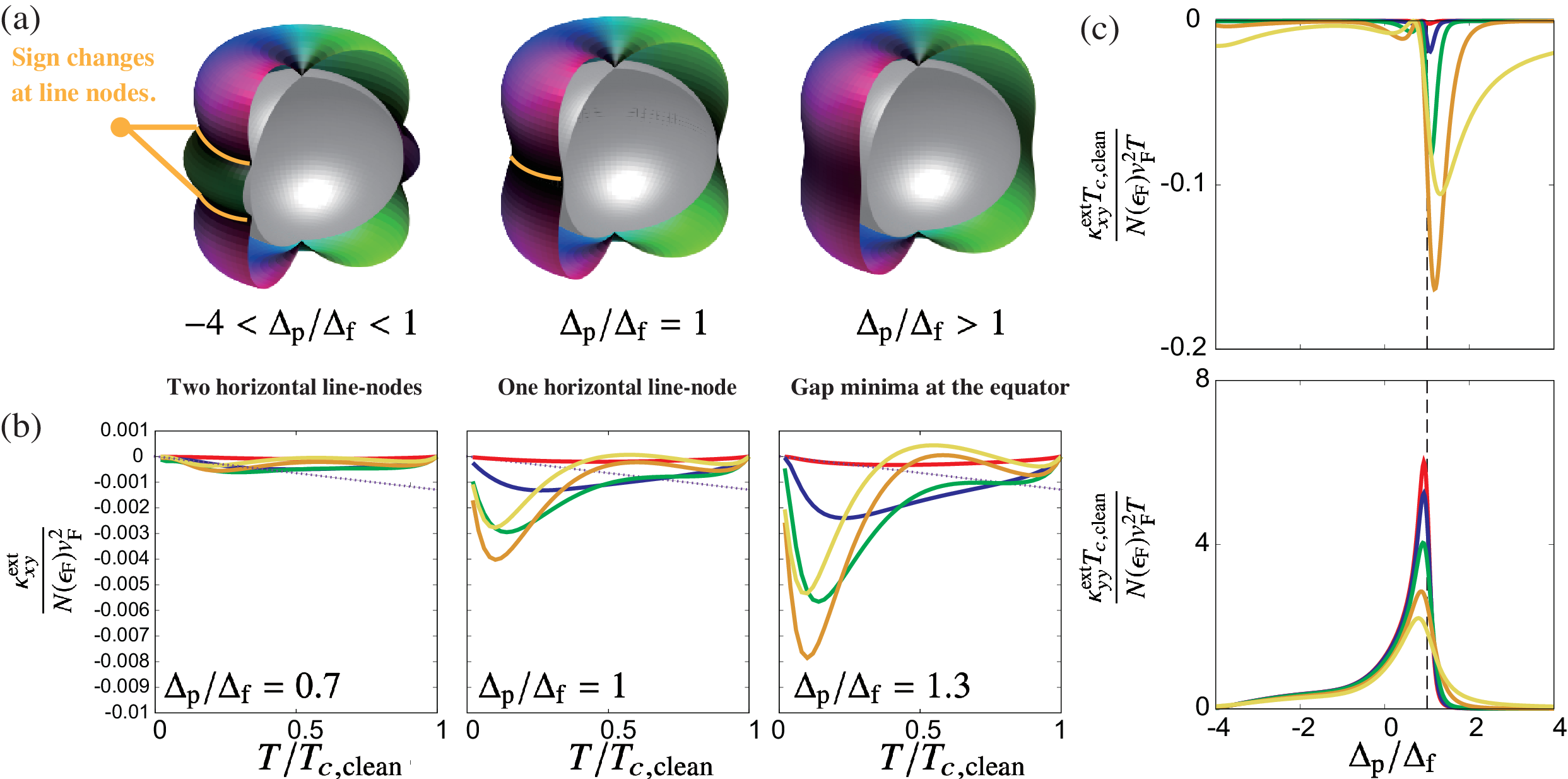}
    \caption{ 
    (a) Schematic picture of the gap structures, (b) temperature dependence of the impurity contribution to the THC, and (c) $\Delta_{\rm p}/\Delta_{\rm f}$ dependence of the thermal conductivity in the mixed pairing states.
    The dotted lines in the panel (b) plot $-\kappa_{xy}^{\rm int}/(N(\epsilon_{\rm F})v_{\rm F}^2)$ using the low-temperature formula with $(k_{\rm F}\xi_0)^{-1}=0.01$, where the minus sign is due to comparing the relative magnitude with $\kappa_{xy}^{\rm ext}$.
    The dashed lines in the panel (c) represent $\Delta_{\rm p}/\Delta_{\rm f}$ where two line nodes merge into one.
    In panels (b-c), we set the scattering rate $\Gamma_{\rm imp}=0.04\pi T_{c,{\rm clean}}$ and the scattering phase-shift $\delta=\frac{\pi}{12}$ (red lines), $\frac{\pi}{6}$ (blue lines), $\frac{\pi}{4}$ (green lines), $\frac{\pi}{3}$ (orange lines), $\frac{5\pi}{12}$ (yellow lines).
    In panel (c), we set the temperature $T=0.02T_{c,{\rm clean}}$.} 
    \label{fig8}
\end{figure*}

Both the $E_{1u}$ chiral $p$-wave and $f$-wave pairings we consider belong to the $E_{1u}$ irreducible representation, and hence these are naturally mixed~\cite{RevModPhys.63.239}.
We denote the chiral $p$ ($f$)-wave order parameter as $\Delta_{\rm p(f)}$ and write the order parameter in the mixed pairing state as,
\begin{eqnarray}
{\bm d}(\bm k_{\rm F})=\Delta_{\rm f}(\hat{k}_x+i\hat{k}_y) \left(5\hat{k}_z^2-1+\frac{\Delta_{\rm p}}{\Delta_{\rm f}}\right)\hat{\bm z}.
\end{eqnarray}
This order parameter in the mixed pairing state is written as Eq.~\eqref{sc_order} with $\eta(\hat{k}_z)=5\hat{k}_z^2-1+\Delta_{\rm p}/\Delta_{\rm f}$.

\begin{figure}[b]
    \includegraphics[width=6cm]{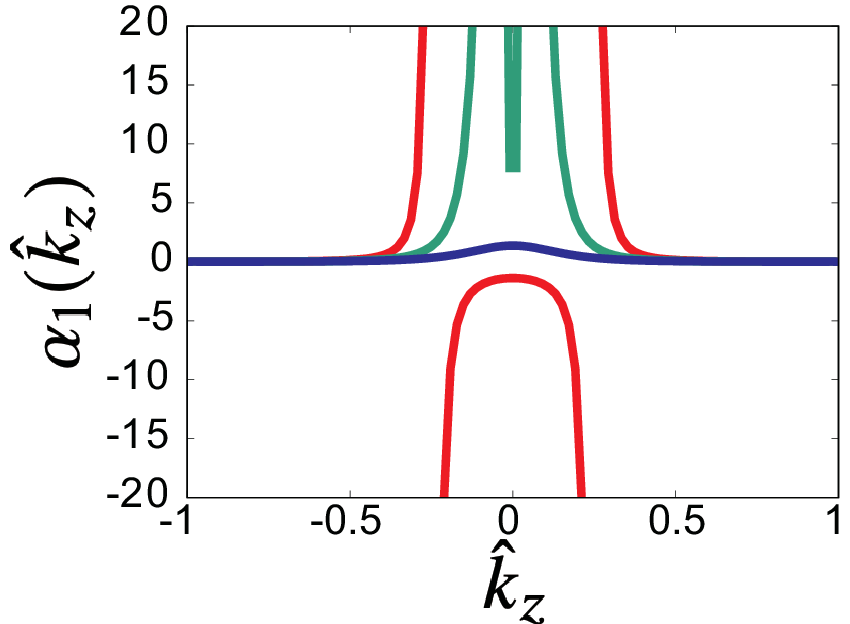}
    \caption{
    The Fermi surface function $\alpha_1(\hat{k}_z)$ in the mixed chiral pairing states [see Eq.~\eqref{alpha}]. 
    The red and green curves represent the calculated results with $\Delta_{\rm p}/\Delta_{\rm f}=0.7,1,1.3$, respectively.
    We set the scattering rate $\Gamma_{\rm imp}=0.04\pi T_{c,\rm clean}$ and the phase-shift $\delta=\frac{\pi}{6}$.}
    \label{fig7}
\end{figure}

The nodal structure of the resulting gap depends on the ratio $\Delta_{\rm p}/\Delta_{\rm f}$.
There are always Weyl nodes at the north and south poles of the Fermi sphere.
In addition, the superconducting gap has two line-nodes when $-4<\Delta_{\rm p}/\Delta_{\rm f}<1$. 
These nodes merge at the equator for $\Delta_{\rm p}/\Delta_{\rm f}=1$. For $\Delta_{\rm p}/\Delta_{\rm f}>1$, the equatorial line node is lifted and replaced by a minimum [see Fig.~\ref{fig8} (a)].

Figure~\ref{fig8} (b) shows that the extrinsic ATHE is enhanced when the two horizontal line nodes merge at the equator.
When the mixed chiral state possesses a single line-node or small line-node gap minima at the equator, the extrinsic ATHE becomes dominant even at low temperatures.
Figure~\ref{fig8} (c) shows that the extrinsic ATHE is sharply suppressed when the line node at the equator splits into two.
As is clear from Fig.~\ref{fig8} (c), the behavior of the ATHE for the gap minima is very similar to the one shown in Fig.~\ref{test}, emphasizing once again that line nodes suppress the extrinsic ATHE, even though they enhance the longitudinal thermal conductivity.

This behavior is also understood from the low-temperature formula for the extrinsic ATHE~[Eq.~\ref{kappa_xy})].
As shown in Fig.~\ref{fig7}, when we introduce the chiral $p$-wave gap to the chiral $f$-wave state with $0<\Delta_{\rm p}/\Delta_{\rm f}<1$, two line nodes get closer to each other.
When the two line nodes approach each other, the positive region of $\alpha_1(\hat{k}_z)$ is enlarged, and its momentum average is enhanced.
As shown in Fig.~\ref{fig7}, when two line nodes merge at the equator, 
$\alpha_1(\hat{k}_z)$ has no sign change and becomes a positive function.
The absence of the sign change at the line nodes enhances the extrinsic ATHE because the contribution of additional quasiparticles in the near-nodal regions (appearing in the denominator of $\alpha_1(\hat{k}_z)$ in Eq.~\eqref{alpha}) is not reduced by the sign change in the numerator on performing the Fermi surface average.

\section{Conclusion}
\label{Sec9}
In this paper, we addressed the relative magnitude of the intrinsic and extrinsic ATHE in WSCs.
The intrinsic ATHE arises from the Berry curvature generated by Weyl points.
The intrinsic ATHE is determined by the structure of the order parameter and the configuration of Weyl points in momentum space. 
On the other hand, the extrinsic contribution arises from the skew scattering of quasiparticles at impurity sites.
Hence, the magnitude of the extrinsic ATHE depends on the detail of parameters for impurities, such as the form of the impurity potential, the scattering rate, and the scattering phase-shift.

The intrinsic ATHE is suppressed by the small factor, $(k_{\rm F}\xi_0)^{-1}$, in superconductors, and thus the extrinsic contribution often dominates the thermal Hall responses.
At the same time, the intrinsic contribution relies on the gapless surface Majorana modes, which ensure $T$-linear behavior in low temperatures regardless of the existence of nodal excitation in bulk~\cite{stone2012gravitational}.
Hence, the intrinsic contribution may dominate the thermal Hall response at low temperatures in clean systems.

We computed the impurity contribution to the THC using the quasiclassical transport theory.
We combined the low-temperature expansion analysis and the numerical calculations to evaluate the effects of the impurity bands and the additional nodal structure on the extrinsic ATHE.
One of our findings is that the variation of the DOS due to impurity bands near the Fermi energy is crucial for the THC.
The impurity contribution to the THC vanishes in both the Born and unitarity limits.
This contribution becomes significant for the intermediate impurity potential strength between these limits. 
The impurity contribution to the THC becomes significant when the tails of the impurity bands touch the Fermi level, amplifying the residual DOS.
We investigated the phase-shift dependence of the DOS and the ATHE and confirmed this behavior.
Furthermore, we derived the formula associating the extrinsic ATHE with the residual DOS [Eq.~\eqref{ATHE_DOS}].

We showed that the additional line-nodal structure significantly reduces the impurity contribution to the THC when line nodes involve the sign change of the order parameter.
For such order parameters, the angular momentum of the Cooper pairs is not the same as the winding number around the nodes.
Consequently, the skew scattering from different parts of the Fermi surface partially compensates. 
In contrast, when the line-nodes are not accompanied by such a sign change, 
quasiparticle excitations around the gap minima significantly enhance the thermal Hall response.

Our results can be applied to any chiral superconductor, even without Weyl nodes.
For an example, we note the chiral $d$-wave pairing, $\Delta(\bm k)\propto  \hat{k}_z (\hat{k}_x\pm i \hat{k}_y)$ among candidate order parameters for Sr$_2$RuO$_4$~\cite{maeno2024still}. 
The chiral $d$-wave order parameter does not generate Weyl points because this compound has quasi-two-dimensional Fermi surfaces, and the Fermi surface is absent at $k_x=k_y=0$.
Although this compound is not a WSC, the chiral $d$-wave superconducting order involves the broken time-reversal and mirror reflection symmetries and allows the ATHE~\cite{yip2016low}.
As described in Sec.~\ref{Sec8}, this chiral $d$-wave order generates a line-node at $k_z=0$ with the sign change of the order parameter, and the extrinsic ATHE is suppressed by the line-nodal excitations.
Hence, the intrinsic contribution will be a substantial part of ATHE in Sr$_2$RuO$_4$, but the detailed balance between the intrinsic and extrinsic contributions depends on the details of the Fermi surface and other material-specific parameters.

For multi-band materials, we need to sum the contributions from each band to obtain the total THC.  
Our result can be used to estimate the contributions from each band.
In principle, the interband effects modify the ATHE, but unless the system is fine-tuned so that the Fermi level lies near degeneracy points, interband contributions are suppressed due to the energy mismatch between the bands at a given wave vector. 
While the detailed analysis of this situation remains a subject for future studies, we expect that our results remain at least qualitatively, and likely quantitatively, valid in most such cases.

Our findings provide clear evidence that in large classes of candidate materials for WSCs, there is likely to be a competition between the intrinsic and extrinsic contributions to the ATHE even at low temperatures.
This suggests that careful (and potentially material-specific) analysis is needed to understand its behavior in the superconducting state. Our work lays the framework and provides the blueprint for such analysis.

\section*{Acknowledgments}
T. Matsuhita thank Y. Masaki, M. Sato, Y. Yanase, Y. Tanaka, J. Ieda, H. Kontani, S. Onari, A. Kobayashi, and R. Tazai for fruitful discussions.
T. Matsushita was supported by a Japan Society for the Promotion of Science (JSPS) Fellowship for Young Scientists, JSPS KAKENHI Grant No.~JP19J20144, No.~24KJ0130, and JST CREST Grant No.~JPMJCR19T2. 
This work was also supported by JST CREST Grant No.~JPMJCR19T5, Japan, and JSPS KAKENHI (Grants No. JP21H01039 and No. JP22H01221).
I. V. was supported in part by grant NSF PHY-1748958 to the Kavli Institute for Theoretical Physics (KITP).

\begin{appendix}

\section{Derivation of Eqs.~\eqref{gEli}, ~\eqref{gEli_ns}, and~\eqref{gEli_vc}}
In this appendix, we derive the anomalous Keldysh QGF, $\delta \check{g}^a$, as the linear response to temperature gradients.
We start from the Eilenberger equation for the nonequilibrium Keldysh QGF, $\delta \check{g}^{\rm K}$,
\begin{align}
\label{K_eq_noneq}
&\left(\underline{M}^{\rm R} \delta \underline{g}^{\rm K}-\delta \underline{g}^{\rm K}\underline{M}^{\rm A}\right)-(\sigma_{\rm imp,eq0}^{\rm R}-\sigma_{\rm imp,eq0}^{\rm A})\delta \underline{g}^{\rm K}\nonumber\\
&+\left( \underline{\sigma}_{\rm imp,eq}^{\rm K}\delta \underline{g}^{\rm A}- \delta\underline{g}^{\rm R}\underline{\sigma}_{\rm imp,eq}^{\rm K}\right)-\left(\delta \underline{\sigma}_{\rm imp}^{\rm R} \underline{g}_{\rm eq}^{\rm K}-\underline{g}_{\rm eq}^{\rm K}\delta \underline{\sigma}_{\rm imp}^{\rm A}\right)\nonumber\\
&-\left( \delta \underline{\sigma}_{\rm imp}^{\rm K}\underline{g}_{\rm eq}^{\rm A}- \underline{g}_{\rm eq}^{\rm R}\delta \underline{\sigma}_{\rm imp}^{\rm K}\right)+(i{\bm v_{\rm F}}\cdot {\bm \nabla} T)\frac{\partial}{\partial T} \underline{g}_{\rm eq}^{\rm K}=0.
\end{align}
Using the anomalous Keldysh QGF and $\delta \underline{\sigma}_{\rm imp}^{a}$, we recast Eq.~(\ref{K_eq_noneq}) into,
\begin{align}
\label{aK_eq_noneq}
&\left(\underline{M}^{\rm R} \delta \underline{g}^{a}-\delta \underline{g}^{a}\underline{M}^{\rm A}\right)-\left( \sigma_{\rm imp,eq0}^{\rm R}-\sigma_{\rm imp,eq0}^{\rm A} \right)\delta\underline{g}^{a}\nonumber\\
&+\left(\underline{g}_{\rm eq}^{\rm R}\delta \underline{\sigma}^{a}- \delta \underline{\sigma}^{a}\underline{g}_{\rm eq}^{\rm A} \right)
-\frac{i\left(\epsilon {\bm v_{\rm F}}\cdot {\bm \nabla} T\right)}{2T^2\cosh^2 \left(\frac{\epsilon}{2T}\right)}
\left( \underline{g}_{\rm eq}^{\rm R}-\underline{g}_{\rm eq}^{\rm A}\right)=0.
\end{align}
$\delta \underline{\sigma}_{\rm imp}^{a}$ is calculated from the $T$-matrix equation,
\begin{align}
\delta \underline{\sigma}^{a}_{\rm imp}=&\Gamma_{\rm imp}\left(\cot \delta+\left\langle\frac{\underline{g}_{\rm eq}^{\rm R}}{\pi}\right\rangle_{\rm FS}\right)^{-1}\nonumber\\
&\times\left\langle\frac{\delta \underline{g}^a}{\pi}\right\rangle_{\rm FS}\left(\cot \delta+\left\langle \frac{\underline{g}_{\rm eq}^{\rm A}}{\pi}\right\rangle_{\rm FS}\right)^{-1}.
\end{align}
The anomalous Keldysh QGF is supplemented by $\underline{g}_{\rm eq}^{\rm R}\delta \underline{g}^{a}+\delta \underline{g}^{a}\underline{g}_{\rm eq}^{\rm A}=0$ from the normalization condition.
Using this condition, we can solve the transport equation [Eq~\eqref{aK_eq_noneq}] and obtain the anomalous Keldysh QGF,
\begin{align}
\label{gEli_ap}
&\delta \underline{g}^{a}=\delta \underline{g}^{a}_{\rm ns}+\delta \underline{g}^{a}_{\rm vc},\\
\label{gEli_ns_ap} 
&\delta \underline{g}^{a}_{\rm ns}=\underline{N}^{\rm R}_{\rm eq}\left( \underline{g}_{\rm eq}^{\rm R}-\underline{g}_{\rm eq}^{\rm A}\right)\left(-\frac{i\left({\epsilon \bm v_{\rm F}}\cdot {\bm \nabla} T\right)}{2T^2\cosh^2\left(\frac{\epsilon}{2T}\right)}\right),\\
\label{gEli_vc_ap} 
&\delta \underline{g}^{a}_{\rm vc}=
\underline{N}^{\rm R}_{\rm eq}\left(\underline{g}_{\rm eq}^{\rm R} \delta \underline{\sigma}_{\rm imp}^{a}- \delta \underline{\sigma}_{\rm imp}^{a}\underline{g}_{\rm eq}^{\rm A} \right),
\end{align}
where
\begin{align}
\label{NR_a@}
\underline{N}^{\rm R}_{\rm eq}=\frac{\left(D^{\rm R}+D^{\rm A}\right)\left(-\frac{\underline{g}^{\rm R}_{\rm eq}}{\pi}\right)+\sigma_{\rm imp,eq0}^{\rm R}-\sigma_{\rm imp,eq0}^{\rm A} }{\left(D^{\rm R}+D^{\rm A} \right)^2+\left( \sigma_{\rm imp,eq0}^{\rm R}-\sigma_{\rm imp,eq0}^{\rm A} \right)^2}.
\end{align}

\section{Derivation of Eqs.~\eqref{kappa_yy} and~\eqref{kappa_xy}}   
\label{sec:app_LTE}
At low temperatures, we can expand the QGF in the frequency $\epsilon$ because the anomalous Keldysh QGF is proportional to the derivative of the equilibrium distribution function, which introduces a small cut-off in the frequency integral in Eq.~\eqref{thermal_current}.
As described in the main text, the change in the spectrum function from equilibrium does not contribute thermal transport. Therefore, we focus on the anomalous Keldysh QGF.

To the leading order in $\epsilon$, the anomalous Keldysh QGF [Eq.~(23)] reduces to,
\begin{align}
    \label{gEli_LT}
    \delta \underline{g}^{a}_{\rm LT}=&\delta \underline{g}^{a}_{\rm ns,LT}+\delta \underline{g}^{a}_{\rm vc,LT},\\
    \label{gEli_ns_LT}
    \delta \underline{g}^{a}_{\rm ns,LT}=&-\frac{\underline{g}^{\rm R}_{\rm eq,LT}}{2\pi D_{\rm LT}}\left( \underline{g}_{\rm eq,LT}^{\rm R}-\underline{g}_{\rm eq,LT}^{\rm A}\right)\nonumber\\
    &\times\left(-\frac{i\left({\epsilon \bm v_{\rm F}}\cdot {\bm \nabla} T\right)}{2T^2\cosh^2\left(\frac{\epsilon}{2T}\right)}\right),\\
    \label{gEli_vc_LT}
    \delta \underline{g}^{a}_{\rm vc,LT}=&-\frac{\underline{g}^{\rm R}_{\rm eq,LT}}{2\pi D_{\rm LT}}\nonumber\\
    &\times\left(\underline{g}_{\rm eq,LT}^{\rm R} \delta \underline{\sigma}_{\rm imp,LT}^{a}- \delta \underline{\sigma}_{\rm imp,LT}^{a}\underline{g}_{\rm eq,LT}^{\rm A} \right),
\end{align}
where the subscript ‘‘${\rm LT}$" represents the zero-frequency limit, $\epsilon=0$.
$\underline{g}^{\rm R,A}_{\rm eq,LT}$ is the equilibrium QGF in this limit, which is given by,
\begin{align}
    &\underline{g}^{\rm R}_{\rm eq,LT}=-\pi\frac{i\gamma \underline{\tau}_z-\underline{\Delta}_{\rm eq}}{D_{\rm LT}},\\
    &\underline{g}^{\rm A}_{\rm eq,LT}=-\pi\frac{-i\gamma \underline{\tau}_z-\underline{\Delta}_{\rm eq}}{D_{\rm LT}},
\end{align}
with $D_{\rm LT}\equiv D^{\rm R}_{\rm LT}=D^{\rm A}_{\rm LT}=\sqrt{\gamma^2+|\Delta_{\rm eq}\hat{k}_{{\rm F}\perp}\eta(\hat{k}_{z})|^2}$.
In the zero-frequency limit, $\sigma_{\rm imp,eq0}$ is purely real and irrelevant for the thermal transport at low-temperatures.
Thus, the equilibrium self-energy expressed as $\underline{\sigma}_{\rm imp,LT}^{\rm R}=-i\gamma \underline{\tau}_z$, where $\sigma_{\rm imp,eq0}$ is neglected.
$\gamma$ is calculated from the $T$-matrix equation,
\begin{align}
    \label{Tmat_zeroen}
    \gamma=\Gamma_{\rm imp}\frac{n_s(0)}{\cot^2\delta +n_s^2(0)},
\end{align}
with the zero-energy DOS,
\begin{align}
    \label{zeroen_DOS}
    n_s(0)=\left\langle \frac{\gamma}{\sqrt{|{\Delta_{\rm eq}\hat{k}_{{\rm F}\perp}\eta(\hat{k}_{z}})|^2+\gamma^2}}\right\rangle_{\rm FS}.
\end{align}

Here, we consider a spatially uniform temperature gradient along the $y$-direction.
We also assumes $\Delta_{\rm eq}\in \mathbb{R}$ without loss of generality.
In this paper, we consider the $d$-vector fixed along the $z$ axis.
In this case, the spin along the $z$ axis is conserved, which allows us to focus on each spin subspace.
Hence, we perform the low-temperature expansion in each spin subspace and drop the spin index.
In the zero-frequency limit, $\delta \underline{\sigma}_{\rm imp}^a$ is calculated from the self-consistent $T$-matrix equation,
\begin{align}
    \label{Tmat_a_LT}
    &\delta \underline{\sigma}_{\rm imp,LT}^a=\frac{\Gamma_{\rm imp}}{(\cot^2 \delta+n_{s}^2(0))^2}\left(\cot \delta +in_s\left(0\right)\underline{\tau}_z\right)\nonumber\\
    &\times\left\langle \frac{\delta \underline{g}^{a}_{\rm ns,LT}+\delta \underline{g}^{a}_{\rm vc,LT}}{\pi} \right\rangle_{\rm FS}
    \left(\cot \delta -in_s\left(0\right)\underline{\tau}_z\right).
\end{align}
The Fermi surface average of the non-selfconsistent contribution, $\delta \underline{g}^{a}_{\rm ns, LT}$ can be straightforwardly performed.
From Eq.~(\ref{gEli_ns_LT}), we obtain,
\begin{align}
    \label{Tmat_a_LT1}
    &\frac{\Gamma_{\rm imp}}{(\cot^2 \delta+n_{s}^2(0))^2}\left(\cot \delta +in_s\left(0\right)\underline{\tau}_z\right)\nonumber\\
    &\times\left\langle \frac{\delta \underline{g}^{a}_{\rm ns,LT}}{\pi} \right\rangle_{\rm FS}
    \left(\cot \delta -in_s\left(0\right)\underline{\tau}_z\right)\nonumber\\
    &=\left(X\underline{\tau}_x+Y\underline{\tau}_y \right)\braket{\alpha_1(\hat{k}_z)}_{\rm FS}\left( -\frac{\Gamma_{\rm imp}\gamma\epsilon v_{\rm F}\Delta_{\rm eq}(-\partial_y T)}{T^2\cosh \left(\frac{\epsilon}{2T}\right)}\right).
\end{align}
From the Eq.~(\ref{Tmat_a_LT}), we make an ansatz for $\delta \underline{\sigma}_{\rm imp,LT}^a$,
\begin{align}
    \label{Tmat_a_LT2}
    \delta \underline{\sigma}_{\rm imp,LT}^a
    =& \left(\tilde{X}\underline{\tau}_x+\tilde{Y}\underline{\tau}_y \right)\braket{\alpha_1(\hat{k}_z)}_{\rm FS}\nn \\
    &\times \left( -\frac{\Gamma_{\rm imp}\gamma\epsilon v_{\rm F}\Delta_{\rm eq}(-\partial_y T)}{T^2\cosh \left(\frac{\epsilon}{2T}\right)}\right).
\end{align}
Using this ansatz, we transform the $T$-matrix equation [Eq.~\eqref{Tmat_a_LT}] into,
\begin{widetext}
\begin{align}
    \label{Tmat_a_LT3}
    \begin{pmatrix}
        1-2\Gamma_{\rm imp}|\Delta_{\rm eq}|^2Y\braket{\alpha_2(\hat{k}_z)}_{\rm FS}&&-2\Gamma_{\rm imp}|\Delta_{\rm eq}|^2X\braket{\alpha_2(\hat{k}_z)}_{\rm FS}\\
        2\Gamma_{\rm imp}|\Delta_{\rm eq}|^2X\braket{\alpha_2(\hat{k}_z)}_{\rm FS}&&1-2\Gamma_{\rm imp}|\Delta_{\rm eq}|^2Y\braket{\alpha_2(\hat{k}_z)}_{\rm FS}
    \end{pmatrix}
    \begin{pmatrix}
        \tilde{X}\\
        \tilde{Y}
    \end{pmatrix}
    =
    \begin{pmatrix}
        X\\
        Y
    \end{pmatrix}.
\end{align}
\end{widetext}
From this matrix equation, we obtain the coefficient for $\delta \underline{\sigma}_{\rm imp,LT}^a$,
\begin{align}
    \tilde{X}&=\frac{X}{\rm Det},\\
    \tilde{Y}&=\frac{Y}{\rm Det}-\frac{\Gamma_{\rm imp}|\Delta_{\rm eq}|^2\braket{\alpha_2(\hat{k}_z)}_{\rm FS}}{8{\rm Det}(\cot^2 \delta+n_{s}^2(0))^2},
\end{align}
where $\rm Det$ represents the determinant of the matrix in Eq.~(\ref{Tmat_a_LT3}),
\begin{align}
    {\rm Det}=&1-\frac{\Gamma_{\rm imp}|\Delta_{\rm eq}|^2(\cot^2 \delta-n_s^2(0))}{(\cot^2 \delta+n_{s}^2(0))^2}\braket{\alpha_2(\hat{k}_z)}_{\rm FS}\nonumber\\
    &+\frac{\Gamma_{\rm imp}^2|\Delta_{\rm eq}|^4}{4(\cot^2 \delta+n_{s}^2(0))^2}\braket{\alpha_2(\hat{k}_z)}_{\rm FS}^2.
\end{align}
We finally obtain the low-temperature formula for thermal conductivities,
\begin{align}
    \frac{\kappa^{\rm ext}_{yy}}{N(\epsilon_{\rm F})v_{\rm F}^2 } \simeq&\frac{\pi^2 T}{6}\gamma^2 \braket{\alpha_0(\hat{k}_z)}_{{\rm FS}}\nonumber\\
    &+\frac{\pi^2 \Gamma_{\rm imp}\gamma^2|\Delta_{\rm eq}|^2T}{3}  \tilde{Y}\braket{\alpha_1(\hat{k}_z)}_{{\rm FS}}^2+\mathcal{O}(T^2),\\
    \frac{\kappa^{\rm ext}_{xy}}{N(\epsilon_{\rm F})v_{\rm F}^2 } \simeq&-\frac{\pi^2 \Gamma_{\rm imp}\gamma^2|\Delta_{\rm eq}|^2T}{3} \tilde{X} \braket{\alpha_1(\hat{k}_z)}_{{\rm FS}}^2+\mathcal{O}(T^2).
\end{align}
In the clean system $\Gamma_{\rm imp}\ll \pi T_c$, the low-temperature formula for the thermal conductivity reduces
to Eqs.~\eqref{kappa_yy} and \eqref{kappa_xy},
\begin{align}
    \frac{\kappa^{\rm ext}_{yy}}{N(\epsilon_{\rm F})v_{\rm F}^2 } \simeq&\frac{\pi^2 T}{6}\gamma^2 \braket{\alpha_0(\hat{k}_z)}_{{\rm FS}}\nonumber\\
    &+\frac{\pi^2 \Gamma_{\rm imp}\gamma^2|\Delta_{\rm eq}|^2T}{3}  Y\braket{\alpha_1(\hat{k}_z)}_{{\rm FS}}^2\nn \\
    &+\mathcal{O}(T^2,\Gamma_{\rm imp}^4),\\
    \label{kappa_xy_ap}
    \frac{\kappa^{\rm ext}_{xy}}{N(\epsilon_{\rm F})v_{\rm F}^2 } \simeq&-\frac{\pi^2 \Gamma_{\rm imp}\gamma^2|\Delta_{\rm eq}|^2T}{3} X \braket{\alpha_1(\hat{k}_z)}_{{\rm FS}}^2\nn \\
    &+\mathcal{O}(T^2,\Gamma_{\rm imp}^4).
\end{align}

\section{Derivation of Eq.~(\ref{ATHE_DOS})}
\label{sec:app32}
We provide the derivation of Eq.~(\ref{ATHE_DOS}) to establish the relationship between the extrinsic ATHE and the change in the DOS due to impurity bands.
We consider the equilibrium $T$-matrix equation and the zero-energy DOS.
Differentiating Eqs.~\eqref{Tmat_zeroen} and \eqref{zeroen_DOS} with the scattering phase-shift, we obtain
\begin{align}
\label{selfen_LT3}
&\frac{\partial \gamma}{\partial (\cot \delta)}= 4\left(Y\frac{\partial n_s(0)}{\partial (\cot \delta)}-X\right),\\
\label{selfen_LT4}
&\frac{\partial n_s(0)}{\partial (\cot \delta)}=|\Delta_{\rm eq}|^2\braket{\alpha_2(\hat{k}_z)}_{\rm FS} \frac{\partial \gamma}{\partial (\cot \delta)},
\end{align}
From Eqs.~(\ref{selfen_LT3}) and (\ref{selfen_LT4}), we obtain
\begin{align}
\label{selfen_LT5}
\frac{\partial n_s(0)}{\partial \delta}\sin^2 \delta=&4\Gamma_{\rm imp}|\Delta_{\rm eq}|^2\braket{\alpha_1(\hat{k}_z)}_{\rm FS}X\nonumber\\
&+\mathcal{O}(\Gamma_{\rm imp}^2).
\end{align}
Comparing Eq.~(\ref{kappa_xy}) to Eq.~(\ref{selfen_LT5}), we find
\begin{align}
\label{ATHE_DOS_ap}
\frac{\kappa^{\rm ext}_{xy}}{N(\epsilon_{\rm F})v_{\rm F}^2 } =&-\frac{\pi^2\Gamma_{\rm imp} \gamma^2T}{12}\frac{\braket{\alpha_1(\hat{k}_z)}_{{\rm FS}}^2}{\braket{\alpha_2(\hat{k}_z)}_{\rm FS}} \frac{\partial n_s(0)}{\partial \delta}\sin^2 \delta \nonumber\\
&+\mathcal{O}(T^2,\Gamma_{\rm imp}^4).
\end{align}
\end{appendix}

\bibliography{ATHE_ref.bib}
\bibliographystyle{apsrev}
\end{document}